\documentclass[12pt,preprint]{aastex}

\usepackage{natbib}

\shorttitle{Charge-Exchange Emission in the Cygnus Loop}
\shortauthors{S. Katsuda et al.}

\begin{document}

\title{Possible Charge-Exchange X-Ray Emission in the Cygnus Loop
  Detected with {\it Suzaku}}

\author{Satoru Katsuda\altaffilmark{1}, Hiroshi
 Tsunemi\altaffilmark{2}, Koji Mori\altaffilmark{3}, 
 Hiroyuki Uchida\altaffilmark{2}, Hiroko Kosugi\altaffilmark{2},
 Masashi Kimura\altaffilmark{2}, Hiroshi Nakajima\altaffilmark{2},
 Satoru Takakura\altaffilmark{2}, Robert Petre\altaffilmark{1}, 
 John W. Hewitt\altaffilmark{1}, \& Hiroya Yamaguchi\altaffilmark{4}
}
%Una Hwang\altaffilmark{1,4}, 

\email{Satoru.Katsuda@nasa.gov}

\altaffiltext{1}{NASA Goddard Space Flight Center, Greenbelt, MD
   20771, U.S.A.} 

\altaffiltext{2}{Department of Earth and Space Science, Graduate School
of Science, Osaka University,\\ 1-1 Machikaneyama, Toyonaka, Osaka,
560-0043, Japan}

\altaffiltext{3}{Department of Applied Physics, Faculty of Engineering,
University of Miyazaki, 1-1 Gakuen Kibana-dai Nishi, Miyazaki, 889-2192,
Japan}

%\altaffiltext{4}{Department of Physics and Astronomy, The Johns Hopkins
%        University, 3400 Charles Street, Baltimore, MD 21218}

\altaffiltext{4}{RIKEN (The Institute of Physical and Chemical
  Research), 2-1 Hirosawa, Wako, Saitama 351-0198} 

\begin{abstract}

X-ray spectroscopic measurements of the Cygnus Loop supernova remnant
indicate that metal abundances throughout most of the remnant's rim
are depleted to $\sim$0.2 times the solar value. However, recent X-ray 
studies have revealed in some narrow regions along the outermost rim
anomalously ``enhanced'' abundances (up to $\sim$1 solar). The reason
for these anomalous abundances is not understood.  Here, we examine
X-ray spectra in annular sectors covering nearly the entire rim of the
Cygnus Loop using {\it Suzaku} (21 pointings) and {\it XMM-Newton} (1 
pointing).  We find that spectra in the ``enhanced'' abundance regions
commonly show a strong emission feature at $\sim$0.7\,keV. This feature is
likely a complex of He-like O K($\gamma$ + $\delta$ + $\epsilon$), 
although other possibilities cannot be fully excluded. The intensity
of this emission relative to He-like O K$\alpha$ appears to be too
high to be explained as thermal emission. This fact, as well as the
spatial concentration of the anomalous abundances in the outermost
rim, leads us to propose an origin from charge-exchange processes between
neutrals and H-like O.  We show that the presence of charge-exchange
emission could lead to the inference of apparently ``enhanced'' metal
abundances using pure thermal emission models.  Accounting for
charge-exchange emission, the actual abundances could be uniformly low
throughout the rim.  The overall abundance depletion remains an open
question.

\end{abstract}

\section{Introduction}

The importance of charge-exchange (CX) emission in X-ray astrophysics
emerged with the discovery of bright X-ray emission  from 
comet Hyakutake (Lisse et al.\ 1996) and the subsequent identification of
its origin from CX processes
between neutral atoms in the comet's atmosphere and highly charged
ions in the solar wind (Cravens et al.\ 1997).  A number of 
comets are now known to have X-ray emission originating from CX
processes (e.g., Cravens 2002).  More recently, the diffuse soft X-ray 
background, which was previously thought to be dominated by thermal
emission from the local hot bubble, turned out to be significantly
contaminated by CX emission from interaction of heliospheric/geocoronal
neutrals and solar wind ions (e.g., Cox 1998; Cravens 2000;
Wargelin et al.\ 2004; Snowden et al.\ 2004; Lallement 2004a; Fujimoto
et al.\ 2007; Smith et al.\ 2007; Koutroumpa et al.\ 2009; Ezoe et
al.\ 2010).  

In principle, CX-induced X-ray emission could be produced at
any astrophysical site where hot plasma interacts with (partially)
neutral gas.  One very promising site is the thin post-shock layer in
supernova remnants (SNRs), since SNR shocks are collisionless and both
unshocked cold neutrals and shocked hot ions can be present just
behind the shock front.  In fact, observational evidence of CX
emission has been obtained as the broad component of H$\alpha$
emission in many SNRs for more than 30 yrs (e.g., Chevalier et al.\
1980; Ghavamian et al.\ 2001).  On the other hand, CX-induced
``X-ray'' emission has not yet been detected firmly.  It has been
suggested only for the SMC SNR 1E0102.2--7219 (E0102) to explain
anomalously high (in the context of thermal emission) intensity ratios
of $\beta$/$\alpha$ and $\gamma$/$\beta$ in O {\small VIII} Ly series
as observed by the {\it XMM-Newton} Reflection Grating Spectrometer
(RGS) (Rasmussen et al.\ 2001).  The authors suggested that CX
processes at the reverse shock contribute significantly to the X-ray
emission.  However, {\it Chandra} High Energy Transmission Grating
(HETG) observations (Flanagan et al.\ 2004) could not confirm the RGS
result: the $\beta$/$\alpha$ ratio was found to be about two thirds of
that measured by the RGS, consistent with that expected in thermal
emission with an electron temperature of $\sim$0.5\,keV.  Although the 
detection of CX-induced X-ray emission from E0102 has been debated,
other SNRs, especially nearby Galactic remnants, are more suitable for
detecting it.  Taking account of the CX emission could be extremely
important for understanding the true properties of the shocked 
plasma, since the undetected presence of CX emission leads to
incorrect plasma parameters (e.g., the metal abundance and the
electron temperature) if we interpret the CX-contaminated X-ray
spectra with pure thermal emission models.   It should be also noted
that CX reactions may be important for understanding magnetic-field
amplification behind the shock front, a topic of continuing interest
in SNR shock physics (Ohira et al.\ 2009; Ohira \& Takahara 2010). 

Unusual metal-abundance regions have been reported at the rim 
of the Cygnus Loop, which is a nearby (540\,pc: Blair et al.\
2005), extended (diameter of $\sim2^{\circ}.8$: Levenson et al.\ 1998) 
SNR: the metal abundances in some of the outermost narrow (a few 
arcminutes, or $\lesssim$0.5\,pc at a distance of 540\,pc) rim regions 
are anomalously ``enhanced'' (up to about 1 solar, hereafter we define
these abundances as ``enhanced'', Katsuda et al.\ 2008ab; Tsunemi et
al.\ 2009; Uchida et al.\ 2009; Kosugi et al.\ 2010), whereas the
metal abundances are typically around 0.2 times the solar value for
most of the rim (hereafter we define these abundances as ``normal'', 
Miyata et al.\ 1999; Leahy 2004; Tsunemi et al.\ 2007; Miyata et al.\
2007; Nemes et al.\ 2008; Zhou et al.\ 2010).  Above and throughout
this paper, we use the solar abundances of Anders \& Grevesse (1989).
The abundance ratios among metals in the ``enhanced'' abundance
regions are consistent with 
solar values within a factor of $\sim$2.  This fact, as well as the
localization of the ``enhancements'' to a few regions along outermost
rim, led the authors to suggest that the ``enhancements'' are not
likely due to contamination by SN ejecta (Katsuda et al.\ 2008a). 
It is believed that the Cygnus Loop is a remnant from a core-collapse
SN and that its forward shock is now hitting the wall of a wind-blown
cavity over considerable fraction of the rim (e.g., McCray \& Snow
1979; Hester et al.\ 1994; Levenson et al.\ 1997).  Therefore,
stellar winds might have altered abundances in part of the
surroundings.  However, stellar winds are expected to be rich in only
lighter elements such as C, N, O, and Ne, as a result of CNO
processing and/or He burning (e.g., Rauscher et al.\ 2002; Murashima
et al.\ 2006), which is inconsistent with the overall abundance
``enhancements''.  It is also unlikely that the ISM abundances are
inhomogeneous over such a small scale (a length scale of several pc),
given that ISM abundances have been reported to be fairly constant in
the local Galaxy.  For example, Cartledge et al.\ (2004; 2006)
revealed that the ISM abundances of a number of sight lines shorter
than 800\,pc are well within the measurement uncertainties, and
concluded that the intrinsic rms scatter of the ISM abundances is
$\sim$10\%.   A similar rms scatter of the ISM abundances was recently
obtained from observations of nearby B stars (Przybilla et al.\ 2008).
Since the Cygnus Loop, whose distance is 540$^{+100}_{-80}$\,pc (Blair
et al.\ 2005), is within the region investigated, the abundance
variation around the rim of the Cygnus Loop is difficult to
understand.  In addition to this problem, the ``normal'' abundances
($\sim$0.2 times the solar value) are also puzzling, given that the
ISM metal abundances around the Cygnus Loop are about half the solar
value (e.g., Cartledge et al.\ 2004).

Noting that the ``enhanced'' abundances are close to the ISM
abundances, Katsuda et al.\ (2008b) and Tsunemi et  al.\ (2009)
suggested that the ``normal'' (i.e., depleted) abundances are only
apparently low due to presence of a nonthermal component (power-law
continuum), which can artificially raise the continuum level and thus
decrease the equivalent widths of the lines.  Indeed, introduction of
a nonthermal component does increase the fitted abundances and yields
slightly better overall spectral fits (the F-test probability of
99\%).  However, the radio flux inferred by the X-ray emission is
unrealistic: it is required to be at least a few orders of magnitude
higher than that observed.  As a result, the authors speculated that
there are indeed two kinds of material with different abundances: one
is the ISM with ``enhanced'' abundances and the other, having the
``normal'' abundances, is the wall of the cavity surrounding the SN
precursor.  There is no good reason, however, why the ISM abundances
should be elevated compared with those of the cavity wall, casting
some doubt on this interpretation.  Although some other possible
explanations of the abundance enhancements and/or depletion, including
dust depletion and resonance-line scattering, have been discussed in
the literature (e.g., Katsuda et al.\ 2008b; Miyata et al.\ 2008;
Uchida et al.\ 2009; Kosugi et al.\ 2010), none provides a
satisfactory explanation.

In this paper, we suggest that the presence of CX emission is
responsible for the abundance ``enhancements'' in the outermost rims
of the Cygnus Loop.  We present a new spatially-resolved X-ray
spectral analysis of nearly the entire rim of the Cygnus Loop using
{\it Suzaku} (21 pointings) and {\it XMM-Newton} (1 pointing).  We
argue that CX-induced X-ray emission is significant in the
``enhanced'' abundance regions and not in the ``normal'' abundance
regions, and that it introduces apparent abundance enhancements when
we apply pure thermal emission models to the CX-contaminated spectra.

\section{Observations and Data Reduction}

\begin{figure}
\includegraphics[angle=0,scale=0.35]{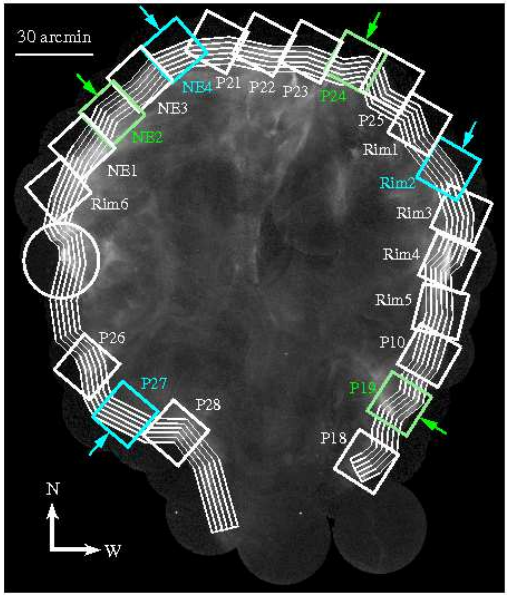}\hspace{1cm}
\caption{Logarithmically-scaled {\it ROSAT} HRI image of the entire
  Cygnus Loop. The image is binned by 5$^{\prime\prime}$ and has been 
  smoothed by a Gaussian kernel of $\sigma = 15^{\prime\prime}$.  The
  FOV analyzed in this paper are shown as boxes for {\it Suzaku} and a
  circle for {\it XMM-Newton}.  For convenience, we label {\it
  Suzaku} FOV according to the object name in the data file.  Spectral 
  extraction regions are indicated as narrow lines along the shock
  front.  XIS spectra extracted from the outermost regions
  in the FOV in cyan (``enhanced'' abundances) and green
  (``normal'' abundances) are shown in Fig.~\ref{fig:cx}.
} 
\label{fig:image}
\end{figure}

We have observed nearly the entire rim region of the Cygnus Loop with
{\it Suzaku}.  In our analysis, we concentrate on the XIS data, since 
hard X-ray emission ($>$10\,keV) from the Cygnus Loop is negligible. 
The fields of view (FOV) of the XIS pointings are shown as white boxes
overlaid on the X-ray mosaic of the Cygnus Loop obtained by 
the {\it ROSAT} HRI (Fig.~\ref{fig:image}).  We also analyze the {\it
XMM-Newton} EPIC data.  {\it XMM-Newton}'s FOV, which is indicated as
a white circle in Fig.~\ref{fig:image}, partly covers the gap in the
{\it Suzaku} observations along the eastern rim.  Information about
the observations is summarized in Table~\ref{tab:obs}. 

We reprocessed the XIS data using the 20090615 version of the CTI
calibration file.  We filtered the reprocessed data using the standard
criteria\footnote{See the {\it Suzaku} data reduction manual which can
be found at   http://heasarc.gsfc.nasa.gov/docs/suzaku/analysis/abc.}  
recommended by the {\it Suzaku}/XIS calibration team.
As for {\it XMM-Newton}, the raw data were processed using version 
8.0.0 of the XMM Science Analysis Software (SAS).  We selected X-ray
events corresponding to patterns
0--12\footnote{http://heasarc.nasa.gov/docs/xmm/abc/}, and removed all 
the events in bad columns listed in Kirsch (2006).  After the 
filtering, the data were vignetting-corrected using the SAS task 
{\tt evigweight}.  The effective exposure times for the various
observations are listed in Table~\ref{tab:obs}.

Background in the XIS arises from two sources: non-X-ray background
(NXB) caused by charged particles and $\gamma$-rays hitting the
detectors, and diffuse X-ray background from the sky.  We generate NXB
spectra suitable for our observations using the 
{\tt xisnxbgen} software (Tawa et al.\ 2008), and subtract them from
the source spectra.  The X-ray background was extensively studied by
Yoshino et al.\ (2009) who modeled it as two thin thermal components
plus one (or two) power-law component(s).  They examined 15 blank-sky
regions in various directions.  Among these, the closest one to the
Cygnus Loop located at ($\ell$, $b$)=(74$^\circ$.0, -8$^\circ$.5) is
``Low Latitude 86--21'' located at ($\ell$, $b$)=(86$^\circ$.0,
-20$^\circ$.8).  In our analysis, we model the X-ray background based
on the spectral parameters for this blank-sky region.  We note that
the X-ray background is much weaker than the Loop emission (which will
be shown in e.g., Fig.~\ref{fig:vpshock}) and that the {\it ROSAT}
all-sky survey shows generally similar backgrounds in these regions
(Snowden, et al.\ 1997).  Therefore, any differences in the background
are negligibly small.  The parameters of the background are taken from
Table~4 in Yoshino et al.\ (2009) except for the total Galactic
hydrogen column density which we modified to
1.7$\times$10$^{21}$cm$^{-2}$ for the direction of the Cygnus Loop
(Leiden/Argentine/Bonn Survey of Galactic HI, Kalberla et al.\ 2005).
For the EPIC background, we use the data set accumulated from blank
sky observations prepared by Read \& Ponman (2003).  We subtract the
blank-sky data from the source after matching the detector
coordinates. 

\section{Comparison of ``Enhanced'' and ``Normal'' Abundance Regions} 

As described in the introduction, there are apparent abundance
inhomogeneities along the rim of the Cygnus Loop: overall, the
abundance is typically 0.2 times the solar value, while parts of the
outermost rim show ``enhanced'' abundances of up to $\sim$1 solar.
Our investigation of the nature of the enhancement starts by
summarizing observational properties of the rim regions.

\subsection{Spectral Properties}

\begin{figure}
\includegraphics[angle=0,scale=0.4]{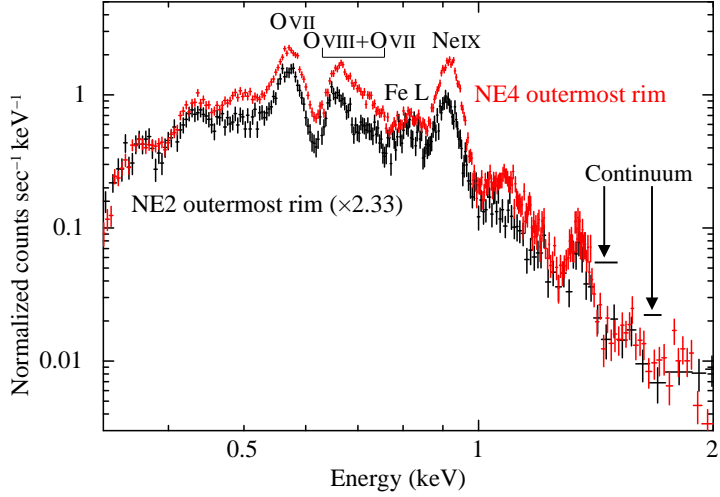}\hspace{1cm}
\caption{Two FI (XIS0+2+3) spectra extracted from the outermost rim
  regions in the NE2 (black) and the NE4 (red).  The
  NE2 spectrum is multiplied by 2.33 so that the count rate 
  in the continuum band (1.4--1.5\,keV and 1.65--1.75\,keV) is
  equalized to that in the NE4 spectrum.  The O and Ne enhancement in
  the NE4 spectrum is readily seen.  
} 
\label{fig:spec_hikaku}
\end{figure}

First, we compare a typical ``enhanced'' abundance spectrum with a
``normal'' one.  To this end, we focus on the XIS data for the NE rim,
because these were obtained in the early phase of the {\it Suzaku}
mission, and have the best photon statistics as well as the best
spectral resolution among all the data listed in Table~\ref{tab:obs}.
These data were previously analyzed by Katsuda et al.\ (2008a), who
divided each FOV into small box regions, and found  ``enhanced''
abundances in the outermost regions in NE3 and NE4.  They also found 
that the parameters are generally constant with azimuth (along the
shock front).  Therefore, to make the comparison with better
statistics, we here extract azimuthally-integrated spectra from narrow 
regions (thickness of 2$^{\prime}$) along the shock front in each FOV
as shown in Fig.~\ref{fig:image}.  Figure~\ref{fig:spec_hikaku} shows
spectra extracted from the outermost regions in NE4 (for the
``enhanced'' abundance region) and NE2 (for the ``normal'' abundance
region).  These spectra are normalized to each other in
the energy bands of 1.4--1.5\,keV and 1.65--1.75\,keV where line
features are not evident.  The differences between the two
are readily seen: K-shell lines from O and Ne are significantly
enhanced in NE4 relative to those in NE2, whereas the complex of Fe
L-shell lines at $\sim$0.82\,keV is comparable in the two spectra. 

\begin{figure}
\plottwo{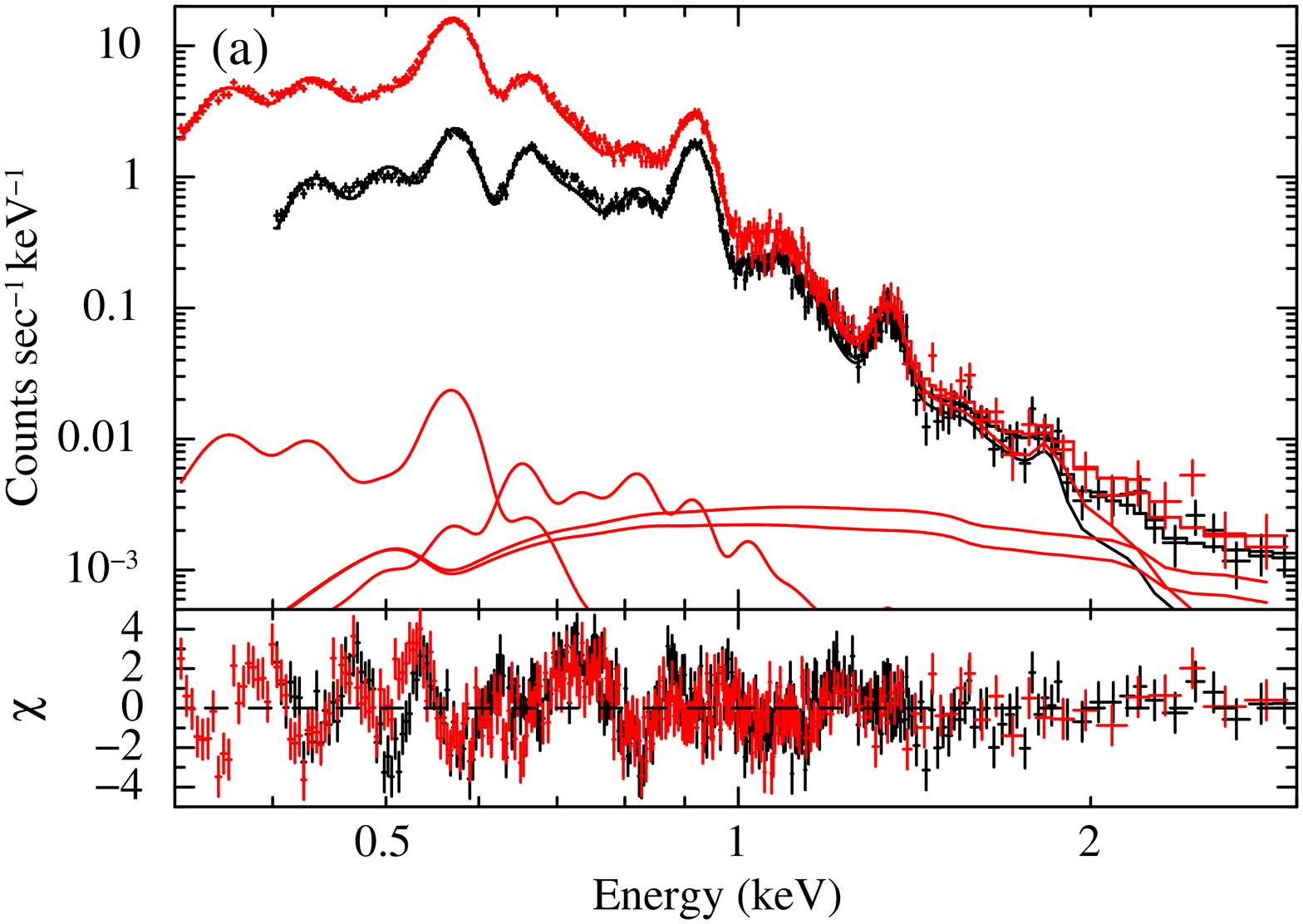}{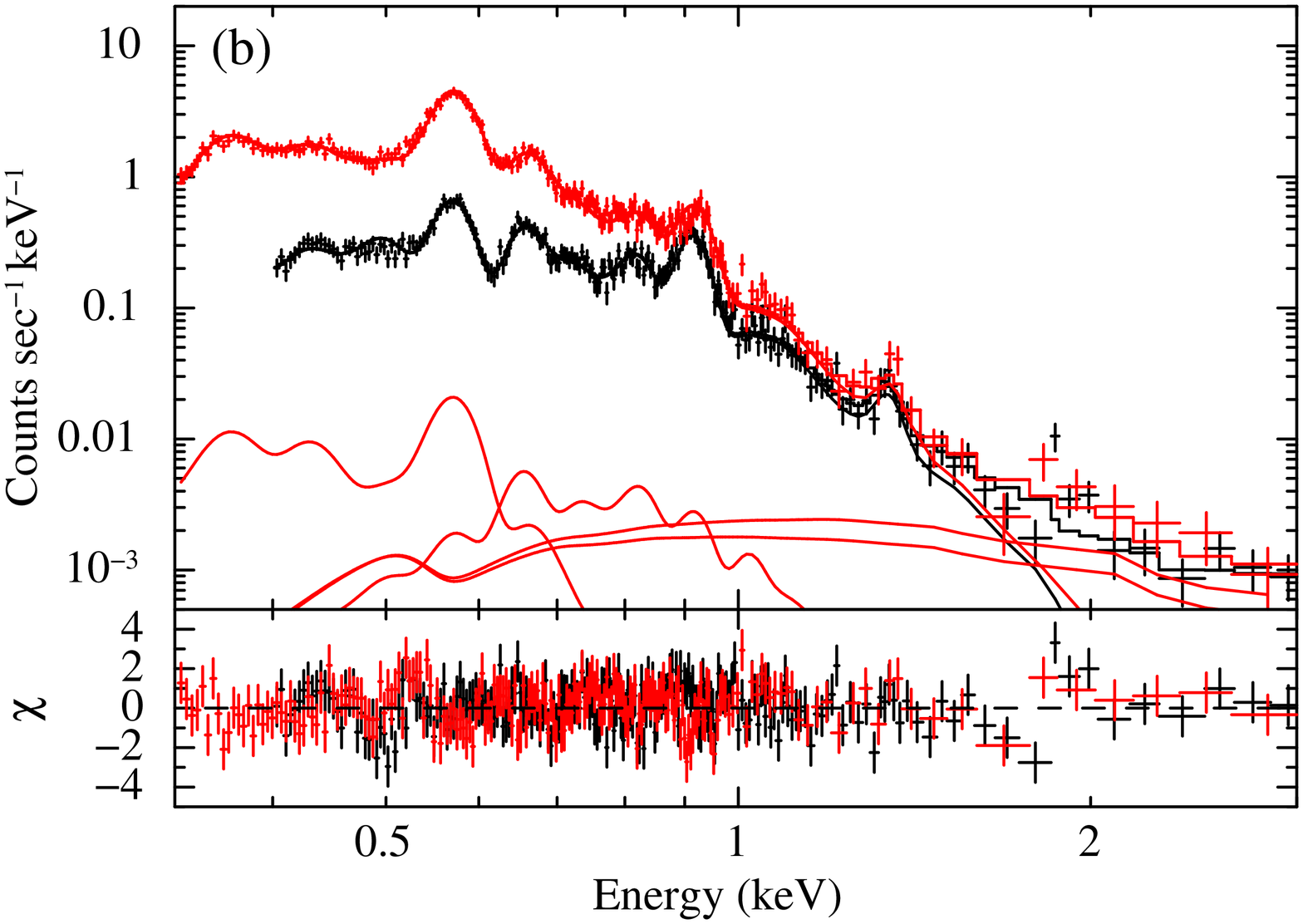}
\caption{(a) XIS spectrum extracted from the NE4's outermost rim along
  with the best-fit single temperature {\tt vpshock} model.  Black and
  red are the FI (XIS0+2+3) and BI (XIS1), respectively.
  Contributions of the X-ray backgrounds (see, text) are separately
  shown only for the BI data.  The residuals are shown in the lower
  panel. (b): Same as left but for NE2. 
}
\label{fig:vpshock}
\end{figure}

To confirm previous fitting results, we fit the two spectra with an
absorbed, single component, plane-parallel shock model [a combination
of the {\tt Tbabs} (Wilms et al.\ 2000) and the {\tt vpshock} model
(NEI version 2.0) (e.g., Borkowski et al.\  2001) in XSPEC v\,12.5.1].
This thermal emission model is the same as that employed in Katsuda et
al.\ (2008b).  We assume the hydrogen column density of the intervening
material to be 3$\times10^{20}$\,cm$^{-2}$ (e.g., Kosugi et al.\
2010). Note that this value is much smaller than the total Galactic
column density of 1.7$\times$10$^{21}$cm$^{-2}$ for the direction of
the Cygnus Loop (see, Section~2) because of the proximity of the Loop.
We freely vary the electron temperature, $kT_\mathrm{e}$, the
ionization timescale, $n_{\rm e}t$, and the volume emission measure 
(VEM $=\int n_\mathrm{e}n_\mathrm{H} dV$, where $n_\mathrm{e}$ and
$n_\mathrm{H}$ are number densities of electrons and protons,
respectively, and $V$ is the X-ray--emitting volume).  Above, $n_{\rm
  e}t$ is the electron density times the elapsed time after shock
heating, and the model uses a range from 0 up to a fitted maximum value.  The
abundances of several elements, whose line emission is prominent in
the spectrum, are treated as free parameters: C, N, O, Ne, Mg, Si
(=S), Fe (=Ni).  We manually adjust the energy scale within the
systematic uncertainties to obtain better fits, by allowing
energy-scale offsets to vary freely for the front illuminated (FI) and
the back illuminated (BI) detectors, respectively.   In this way, we
fit the spectra in energy ranges of 0.4--3.0\,keV for FI and
0.33--3.0\,keV for BI.  Figure~\ref{fig:vpshock} shows the best-fit
model along with the data points.  The best-fit parameters are
summarized in Table~\ref{tab:vpshock_param}.  We confirm that the 
absolute (relative to H) abundances are enhanced in NE4 compared with
those in NE2, and that the abundance ratios of C/Fe, N/Fe, O/Fe, and
Ne/Fe are $\sim$2--4 times higher in NE4 than those in NE2.
We also confirm that the plasma is in the non-equilibrium ionization
(NEI) condition ($n_\mathrm{e}t < 10^{12}$\,cm$^{-3}$\,sec$^{-1}$).
Dividing the ionization timescale by the time elapsed after shock
heating, which is roughly inferred from the distance of
$\sim1\times10^{18}$\,cm (i.e., 2$^\prime$ at a distance of 540\,pc)
divided by the fluid velocity of $\sim$200\,km\,sec$^{-1}$ 
(e.g., Salvesen et al.\ 2009), we estimate the post-shock electron
densities in NE4 and NE2 to be $\sim$1\,cm$^{-3}$ and
$\sim$6\,cm$^{-3}$, respectively.  Assuming strong shocks, we can
estimate the pre-shock ambient densities in NE4 and NE2 to be
$\sim$0.25\,cm$^{-3}$ and $\sim$1.5\,cm$^{-3}$, respectively.  These
values generally agree with previous estimates (e.g., Raymond et al.\
2003 for NE4; Hester et al.\ 1994 for NE2), showing the robustness of
the ionization timescales derived in these outermost rim regions
from the very shock fronts to 2$^{\prime}$ inside the shocks.

We notice that the fit quality for the NE4 spectrum is far from 
formally acceptable (reduced $\chi^2$ of $\sim$2.2), while the fit
quality for the NE2 spectrum is much better.  The NE4 spectrum in
Fig.~\ref{fig:vpshock} (a) shows 
apparent excess emission at $\sim$0.7\,keV relative to the best-fit
model.  This discrepancy was already noticed in the 
{\it Chandra} analysis of the NE rim (Katsuda et al.\ 2008b), although
it was not so evident due to poorer statistics.  We therefore re-fit
the NE4 spectrum using the same model, but ignoring the energy band
0.68--0.76\,keV.  Table~\ref{tab:vpshock_param} (the 3rd column)
summarizes the best-fit parameters.  The fit quality is
significantly improved over the original one (the reduced-$\chi^2$ is 
reduced from 2.2 to 1.6).  Both the electron temperature and
the abundances significantly decrease from the original best-fit
values to values more consistent with those in NE2 (the 2nd column in
Table~\ref{tab:vpshock_param}).  These changes are interpreted as
follows.  In the original (entire energy band) fitting,  
the He-like O K-shell lines ($n\geq4 \to n=1$) which dominate around
0.7\,keV are forced to be as strong as possible to match excess
emission in the 0.68--0.76\,keV band.  The way these lines are made
stronger is by an increase in the electron temperature as well as the
O abundance.  When this energy band is disregarded in the fit, the
electron temperature decreases from $\sim$0.3\,keV to $\sim$0.2\,keV.
The reduced temperature in turn allows a reduced O abundance, since
the emissivities of O K-shell lines peak at $kT_{\mathrm
  e}\sim$0.2\,keV. Other metal abundances are also reduced, with
the net result that the relative abundances among metals are 
maintained.  Therefore, we conclude that the 0.68--0.76\,keV band
plays an important role in determining absolute abundances in the
``enhanced'' abundance spectrum when using the {\it Suzaku} XIS or
any other CCD spectrometers.  For the ``normal'' abundance spectrum
(e.g., the NE2 spectrum in Fig.~\ref{fig:vpshock} (b)), ignoring the 
0.68--0.76\,keV band does not change the fit quality or the best-fit
parameters.

In order to compare the two spectra in more detail, we next fit them
with a phenomenological model consisting of an absorbed,
bremsstrahlung continuum plus a number of Gaussians for line emission
from C, N, O, Ne, Mg, Si, and Fe.  Again, we assume the hydrogen
column density to be 3$\times10^{20}$\,cm$^{-2}$.  We include eight
Gaussians for the prominent line features at $\sim$0.35\,keV (C
Ly$\alpha$), $\sim$0.45\,keV (C Ly$\beta$ $+$ N He$\alpha$), 
$\sim$0.5\,keV (N He$\beta$ $+$ N Ly$\alpha$), $\sim$0.57\,keV (O 
He$\alpha$), $\sim$0.91\,keV (Ne He$\alpha$), 1.35\,keV (Mg
He$\alpha$), $\sim$1.6\,keV (Mg He$\beta$), and $\sim$1.85\,keV (Si
He$\alpha$).  The center energies of these lines are allowed to vary  
freely.  In addition to these, we introduce nine Gaussians to
represent lines that could contribute in the crowded 0.7--1.2\,keV
band.  These are O Ly$\alpha$ at 0.654\,keV, O He$\beta$ at
0.666\,keV, an Fe {\small XVII} L-shell complex from the 3s$\to$2p
transitions at 0.726\,keV, O Ly$\beta$ at 0.775\,keV, an Fe {\small
XVII} L-shell complex from the 3d$\to$2p transitions at 0.822\,keV, Ne
Ly$\alpha$ at 1.022\,keV, Ne He$\beta$ at 1.074\,keV, Ne He$\gamma$ at
1.127\,keV, and Ne Ly$\beta$ at 1.210\,keV.  The line center energies
of these nine Gaussians are fixed at their theoretically expected
values (APED: Smith et al.\ 2001).  Based on the most recent NEI
plasma code (Borkowski et al.\ 2001; the augmented NEI version 2.0
that includes inner-shell processes), the 
intensity ratio between overall Fe L-shell 3s$\to$2p and 3d$\to$2p
lines is fairly constant at unity.  Therefore, we assume the 
normalization of the Gaussian for the 3s$\to$2p transition lines to be
the same as that for the 3d$\to$2p transition lines.  We manually
adjust the energy scale within the systematic uncertainty.  In this
way, we fit the two spectra shown in Fig.~\ref{fig:spec_hikaku}.  As
expected from the spectral fitting in the previous paragraphs, we find
excess emission from the best-fit model at $\sim$0.7\,keV for the
``enhanced'' abundance spectrum.  Addition of a Gaussian at
$\sim$0.7\,keV gives a satisfactory fit; the statistical significance
of introducing the line is greater than 99.9\% for both  
spectra (reduced-$\chi^2$ values are reduced from 1297 to 601 for NE4
and from 440 to 404 for NE2). 

\begin{figure}
\includegraphics[angle=0,scale=0.6]{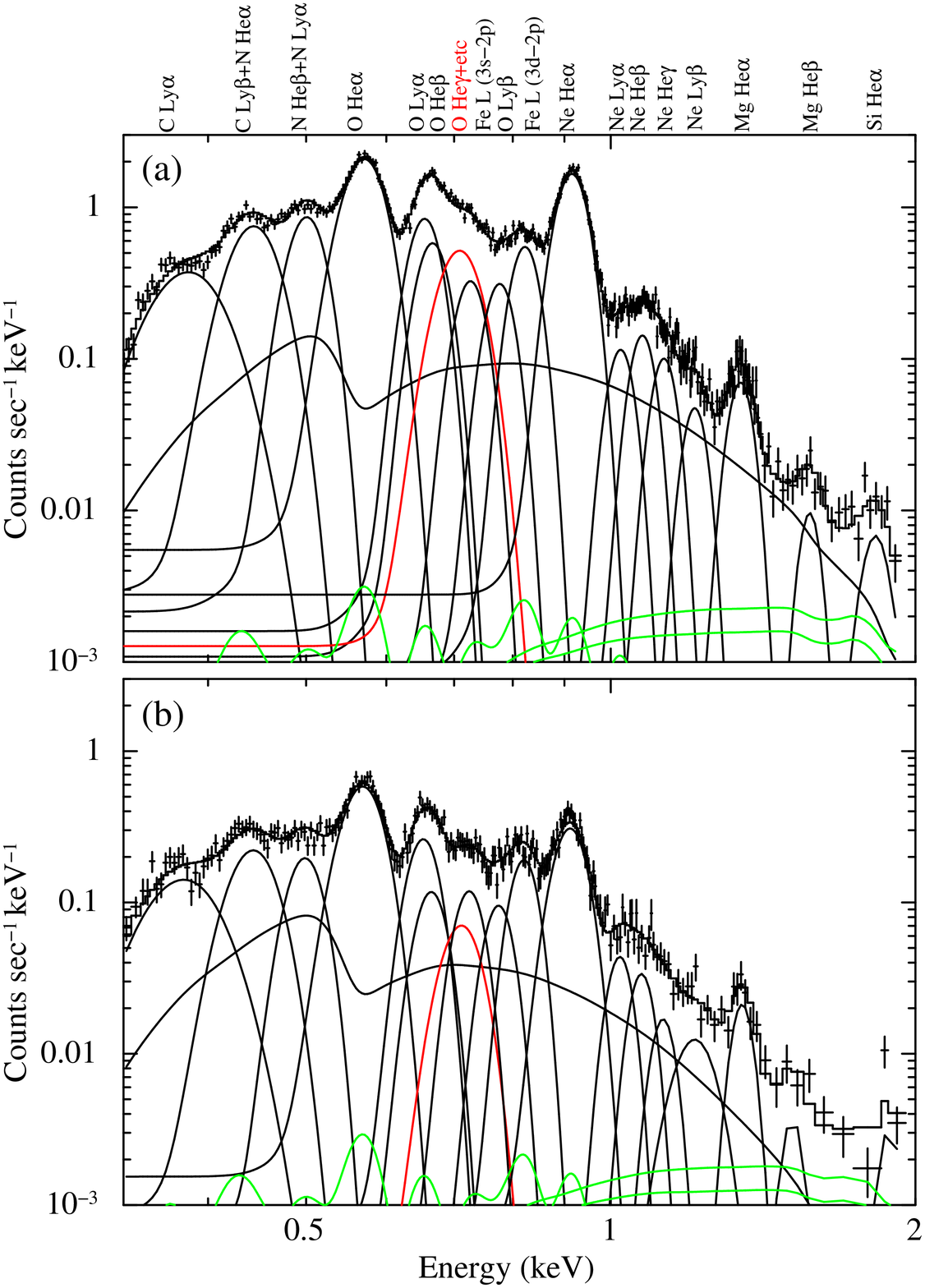}\hspace{1cm}
%\plottwo{f4a.eps}{f4b.eps}
\caption{(a) NXB-subtracted FI (XIS0+2+3) spectrum from the NE4's
  outermost rim (FI only) along with the best-fit model consisting of
  a bremsstrahlung plus 18 Gaussians.  Line identifications are shown
  on top of the panel.  Green curves show individual contributions of
  the X-ray background.  (b) Same as above but for the NE2's outermost
  rim.
}
\label{fig:bremss_gauss}
\end{figure}

The electron temperatures of the bremsstrahlung components are 
0.16$\pm$0.01\,keV and 0.26$\pm$0.01\,keV for NE2 and
NE4, respectively.  Although we do not consider radiative
recombination continuum emission or two-photon decay continuum
emission, the electron temperatures derived are reasonable 
for the rim of the Cygnus Loop.  Table~\ref{tab:bremss_gauss}
summarizes details of this model.  The line center energies listed in
the table include the effects of energy-scale shifts determined by the
fittings.  Figure~\ref{fig:bremss_gauss} presents the best-fit models.
We find that the line intensities in NE4 are generally $\sim$3 times
higher than those in NE2, whereas the line at $\sim$0.7\,keV indicated
as a red line in Fig.~\ref{fig:bremss_gauss} is $\sim$7 times stronger
in NE4 than in  NE2.  Its nonzero width suggests that it is a complex
of lines, although we cannot fully eliminate the possibility that the
broadening might be due to calibration effects.

%In summary, we find a much stronger line complex at $\sim$0.7\,keV in the 
%``enhanced'' abundance spectrum extracted from the NE rim relative to
%the ``normal'' abundance spectrum.  This line complex cannot be
%explained by the current NEI plasma model, as was hinted in Katsuda et
%al.\ (2008b).  The absolute abundances obtained depend strongly on
%whether or not we use the energy band including the $\sim$0.7\,keV 
%complex in our spectral fitting.

\subsection{Spatial Properties}

\begin{figure}
\includegraphics[angle=0,scale=0.65]{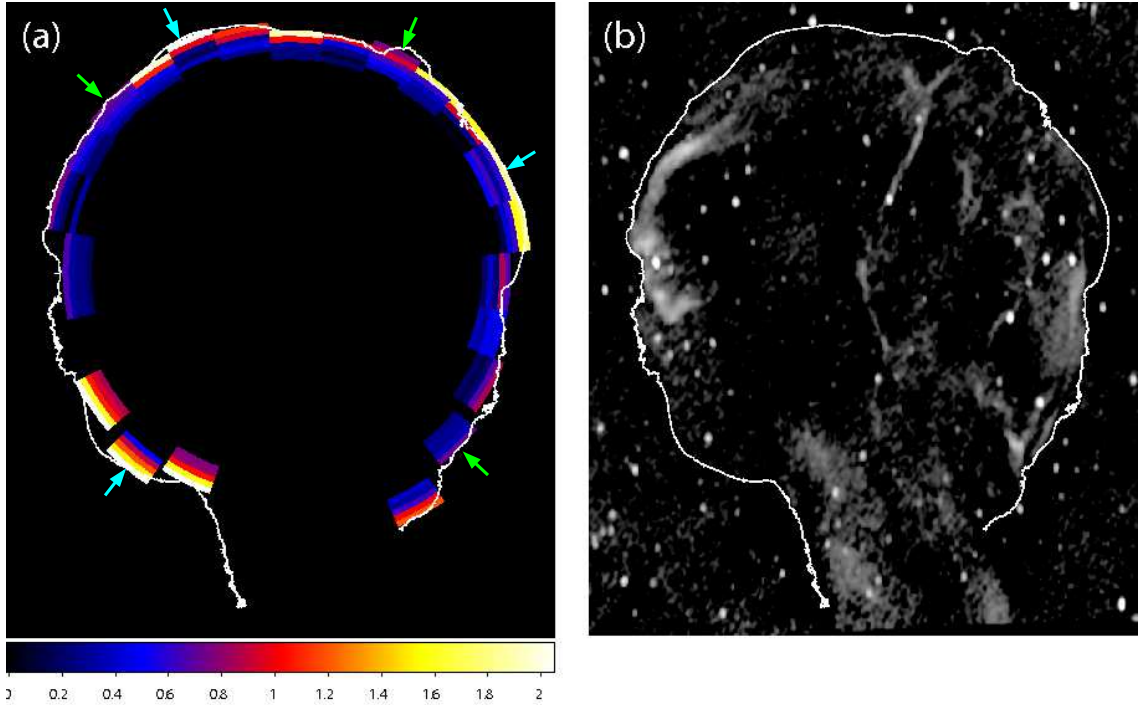}\hspace{1cm}
\caption{(a) Line-ratio map of the $\sim$0.7\,keV emission to the Fe
  L-shell complex at $\sim$0.82\,keV.  The X-ray boundary of the
  Cygnus Loop is shown as a white line.  Arrows indicate the regions
  where we show the XIS spectra in Fig.~\ref{fig:cx}.
  (b) Logarithmically-scaled radio (325 MHz) surface brightness map
  obtained by the WENSS team (Rengelink et al.\ 1997), with the X-ray  
  boundary of the Cygnus Loop as in Fig.~\ref{fig:map} (a).
}
\label{fig:map}
\end{figure}

Before proceeding to an investigation of the origin of the
$\sim$0.7\,keV feature, we briefly summarize the spatial properties of 
the $\sim$0.7\,keV emission.  To reveal the distribution of the
$\sim$0.7\,keV feature, we fit all the spectra for all the regions in 
Fig.~\ref{fig:image} using the same phenomenological model as we
employed in Section~3, a bremsstrahlung plus 18 Gaussians.
We show example best-fit values for the outermost rim regions in P27,
P19, Rim2, and P24 in Table~\ref{tab:bremss_gauss}.  After 
deriving individual line intensities in each region, we calculate the 
line ratio between the $\sim$0.7\,keV emission and the Fe L-shell
complex at $\sim$0.82\,keV.  This ratio should be a good tracer of the
relative strength of the $\sim$0.7\,keV emission, given that the
abundance (and intensity) of Fe (L-shell complex) is relatively
constant in the rim regions (see, e.g., Fig.~\ref{fig:spec_hikaku}).  
Figure~\ref{fig:map} (a) shows this line-ratio map, with the X-ray
boundary of the Cygnus Loop in white.  The map clearly shows that the
$\sim$0.7\,keV emission is enhanced exclusively within a narrow
($\lesssim4^{\prime}$) layer behind the shock.

Also notable from Fig.~\ref{fig:map} (a) is the azimuthal variation.
The $\sim$0.7\,keV emission is evident only at position angles of
0$^\circ$--40$^\circ$ (N--NE), 110$^\circ$--160$^\circ$ (SE), and
270$^\circ$--330$^\circ$ (northwest: NW) measured from north over 
east.  These are the same locations as where the 
``enhanced'' abundances have been reported so far (N--NE by Katsuda et 
al.\ 2008ab and Uchida et al.\ 2009; SE by Tsunemi et al.\ 2009 and 
Kosugi et al.\ 2010).  In addition, preliminary spectral analysis of
the NW region shows ``enhanced'' abundances at the outermost regions
compared with those in the inner regions (Takakura et al.\ in
preparation).  Therefore, it appears that the ``enhanced'' abundance
regions commonly show the strong $\sim$0.7\,keV emission.  We further
note that a map of the intensity ratio of Ne He$\alpha$ to the Fe
L-shell complex at $\sim$0.82\,keV is quite similar to
Fig.~\ref{fig:map} (a), and that the intensity-ratio enhancements seen
in part of the region (which corresponds to the ``enhanced''
abundance regions) are statistically significant.

Figure~\ref{fig:map} (b) shows a 325 MHz radio image of the Cygnus
Loop from the The Westerbork Northern Sky Survey (WENSS) (Rengelink et
al.\ 1997) covering the same sky region of Fig.~\ref{fig:map} (a).
Looking at Fig.~\ref{fig:map} (a) and (b), we notice that the regions
of strong $\sim$0.7\,keV emission are generally anti-correlated with
radio-bright regions.

\section{Discussion}

\subsection{Identification of Excess 0.7\,keV Emission in the
  ``Enhanced'' Abundance Regions}

In this section, we discuss the origin of the $\sim$0.7\,keV emission
feature.  The center energy of the $\sim$0.7\,keV feature suggests
that it is most likely either O K-shell emission or Fe {\small
XVII}/{\small XVI} L-shell emission.  We first consider Fe {\small 
XVII}.  In the X-ray band, there are two well-known strong Fe 
L-shell complexes at $\sim$0.73\,keV and $\sim$0.82\,keV, which mostly
arise from the 3s$\to$2p and the 3d$\to$2p transitions in Fe {\small
  XVII} ions, respectively.  While these two complexes are included in
the phenomenological model above, the intensities of the two were set
equal, based on the most up-to-date plasma code 
(Borkowski et al.\ 2001).  This assumption leaves room for the
3s$\to$2p complex to contribute to the $\sim$0.7\,keV complex, given
the moderate spectral resolution of the XIS (FWHM$\sim$60\,eV at
1\,keV).  However, the energy of the 3s$\to$2p complex is
$\sim$0.73\,keV, making it unlikely to be the dominant source of the
$\sim$0.7\,keV emission, the center energy of which is
0.705$\pm0.002_\mathrm{stat}\pm0.005_\mathrm{sys}$\,keV (the systematic 
uncertainty is from Koyama et al.\ 2007).  In addition, if
we attribute the entire intensity of the $\sim$0.7\,keV emission
to the 3s$\to$2p complex, the intensity ratio of 3s$\to$2p/3d$\to$2p
would be $\sim$3.5.  Such a high ratio has never been observed in an  
astrophysical object.  Although astrophysical sources including
stellar coronae, SNRs, and elliptical galaxies do tend to show higher 
3s$\to$2p/3d$\to$2p ratios than the predicted value (Doron \& Behar
2002 and references therein), the observed ratio ranges from 1 to
2.  Our own analyses at {\it XMM-Newton} RGS observation of the LMC SNRs
DEM~L71, N103B, N132D, N49, and N63A reveal that the
3s$\to$2p/3d$\to$2p intensity ratio in each is near or less than
unity.  Therefore, we rule out 3s$\to$2p emission from Fe {\small
  XVII} as the origin of the $\sim$0.7\,keV feature. 

Turning to Fe {\small XVI} L-shell lines, Graf et al.\ (2009) recently
reported relative intensities, using University of California /
Lawrence Livermore National Laboratory's EBIT-I.  In the energy range
of 0.69--0.86\,keV, the authors measured ten Fe {\small XVI} lines
originating from innershell ionization processes.
Among them, the strongest one is located at $\sim$0.82\,keV, whereas 
four relatively weak lines are clustered around 0.7\,keV.  The total
intensity of the lines around 0.7\,keV is weaker than that
around 0.82\,keV.  This measurement is inconsistent with our measured line 
ratio: the $\sim$0.7\,keV emission is $\sim$2.5 times stronger than 
the $\sim$0.82\,keV emission (which should include both the Fe {\small 
XVI} L-shell lines and the Fe {\small XVII} 3d$\to$2p complex).  It is
therefore unlikely that the Fe {\small XVI} L-shell lines dominate the
$\sim$0.7\,keV emission, although we cannot fully exclude this
possibility given that these lines will incidentally be stronger in an
NEI plasma than in a collisional ionization equilibrium plasma.

Finally, we consider the possibility that O {\small VII} K-shell lines
($n\geq4 \to n=1$) produce the $\sim$0.7\,keV emission.  In this case,
we again run into a line ratio problem for a thermal spectrum:  the
intensity ratio of the $\sim$0.7\,keV emission to O He$\alpha$ is
measured to be $\sim$0.06 (see, the parameters in
Table~\ref{tab:bremss_gauss}), whereas the O He($\gamma+\delta$) to O
He$\alpha$ ratio is expected to be $\sim$0.02 at
$kT_\mathrm{e}=0.2$\,keV based on the NEI code (Borkowski et al.\
2001).  Therefore, it is also unlikely that thermally-emitted O
{\small VII} K-shell lines are responsible for the $\sim$0.7\,keV
emission.

\subsection{Charge-Exchange Emission}

We suggest here that a pure NEI plasma model is not appropriate for
reproducing the $\sim$0.7\,keV emission in the ``enhanced'' abundance
region, and that an additional emission mechanism is required.  One
possible emission mechanism, suggested from the interpretation of the {\it
  XMM-Newton} RGS spectrum of the SMC remnant E0102 (Rasmussen et 
al.\ 2001), is CX-induced emission between neutrals and ions at the
forward shock of the Cygnus Loop.

CX emission would play an important role only in a thin region 
immediately behind the forward shock in a SNR, because cold neutrals
cannot survive far behind the shock.  Taking account of line-of-sight
effects, CX emission if present could only be detected at the
periphery of a SNR.  This is exactly what the line ratio map in
Fig.~\ref{fig:map} (a) shows.

In general, the presence of neutral hydrogen is essential for
CX emission to be significant.  The presence of neutral hydrogen 
in the post-shock region can be tested by the presence/absence
of non-radiative H$\alpha$ filaments.  Such filaments are found
around nearly the entire perimeter of the Cygnus Loop (Levenson et al.\ 
1998), clearly indicating the presence of neutral hydrogen in the
post-shock regions.  One H$\alpha$ filament along the NE
rim was extensively studied by Ghavamian et al.\ (2001), who modeled
the broad and narrow components of the H$\alpha$ and H$\beta$ emission
with their shock model to measure the degree of electron-ion
temperature equilibration.  In their model, the neutral fraction in
the ambient medium is also a measurable parameter, although not
a sensitive one.  They examined three cases of neutral
fractions, i.e., 0.1, 0.5, and 0.9, to model the broad-to-narrow
line ratios, and found that the first case is slightly outside the
range of the observational uncertainties whereas the latter two are
consistent with the observation.  Thus, the preferred range for the
neutral fraction is larger than 0.1 in this region.

With non-radiative H$\alpha$ filaments around nearly the entire
periphery, and a non-zero neutral fraction, it might be expected that
the CX emission should be closely correlated with the non-radiative
filaments.  However, it does not seem to be the case.  One possible
explanation for distribution of CX emission is suggested by 
the radio map (Fig.~\ref{fig:map}).   The regions of the rim from
which the hypothetical CX emission is not detected generally
correspond to those that are radio bright.  The radio emission is
likely produced by accelerated cosmic-rays (electrons), since the
spectral shape of the radio emission is well described by a power-law
continuum (e.g., Uyaniker et al.\ 2004).  Therefore, the
anti-correlation can be qualitatively understood in a view that the
cosmic-ray shock precursor more effectively pre-ionizes the medium,
and thus reduces the CX emission below a detectable level.  It is also
interesting to note that the radio-bright (or CX-absent) regions
correspond to radiative 
filaments (Levenson et al.\ 1998).  This correspondence suggests that
the ambient densities there are higher and the shock velocities are
lower than elsewhere along the rim.  According to Lallement (2004b),
slower  shocks produce less CX emission, providing another possible 
explanation for the observed azimuthal variation of the CX emission.
It may be also possible that thermal emission overwhelms CX emission
in these radio/optical-bright regions.

All of these observations support the possibility of CX emission in
the ``enhanced'' abundance regions.  Furthermore, a major benefit of
this scenario is that the introduction of CX emission resolves the
mysterious abundance ``enhancements'' in the outermost rim regions, by
reducing the ``enhanced'' abundances to consistency with the
``normal'' abundances found everywhere else around the rim.  We
therefore suggest that CX emission is contributing significantly to
the X-ray spectrum in the ``enhanced'' abundance regions.

We note that if this suggestion is to hold, CX emission other than the
$\sim$0.7\,keV complex is expected to be present.  As pointed out in
Section~3.2, the similarity of a map of the intensity  
ratio of Ne He$\alpha$ to the Fe L-shell complex to Fig.~\ref{fig:map}
suggests that Ne He$\alpha$ is also contaminated by CX-induced lines.
In addition, under the condition of the electron temperature of
$\sim$0.2\,keV, C is likely fully ionized and thus H-like C emission
due to CX processes would be also significant.  CX strongly enhances
high principal number Ly emission such as C Ly$\gamma$ at 
$\sim$0.46\,keV (Wargelin et al.\ 2008).  This emission can
contaminate the C Ly$\beta+$N 
He$\alpha$ blend around 440\,eV (see, Table~\ref{tab:bremss_gauss}).
Therefore, the line intensity ratio of this blend to C Ly$\alpha$ may
be enhanced where we find indications of CX.  However, such a trend
can not be detected as shown in Table~\ref{tab:bremss_gauss}.  
A definitive identification of the possible CX lines would require 
higher spectral resolution observations which will be soon available
by the Soft X-ray Spectrometer (SXS) with the energy resolution of
4--7\,eV and the energy calibration of 1--2\,eV (Mitsuda et al.\ 2010)
onboard the {\it Astro-H} satellite (Takahashi et al.\ 2010).

\subsection{Supporting Evidence for CX Emission}

\subsubsection{Spectral Modeling of CX Emission}

In order to further study the properties of the proposed CX emission,
we model CX-contaminated spectra by a combination of a number 
of Gaussians for the CX emission and a plane parallel shock model (the 
{\tt vpshock} model; Borkowski et al.\ 2001) for the thermal emission.
We examine three spectra extracted from the outermost rim regions in 
NE4, P27, and Rim2 where the possible CX emission is fairly strong 
(see, Fig.~\ref{fig:map}).
The challenge is the difficulty in separating CX emission from thermal 
emission in CCD spectra.  Detailed modeling considering the effects of 
CX on the ionization balance or evolution of the electron temperature is 
far beyond the scope of this paper.  We here crudely assume that the 
shape of the thermal emission in this region is similar to that in its 
surroundings, so that we can estimate the thermal emission in the 
CX-contaminated spectra.  Since the thermal spectra are expected to be 
more constant with azimuth than with the distance from the shock front, 
we use the thermal spectrum extracted from the outermost rim in
observation fields where we 
do not find indications of CX emission.  As a template of thermal
emission, we take the NE2 spectrum shown in
Figs.~\ref{fig:spec_hikaku} and \ref{fig:vpshock}.  The best-fit
spectral parameters are listed in Table~\ref{tab:vpshock_param}.  It
should be noted that recombination rates of ions increase due to CX
processes and thus plasmas showing CX emission may be in lower
ionization states than those showing pure thermal emission.  However,
it is quite difficult to estimate this effect quantitatively.  We thus
evaluate systematic uncertainties by examining two ionization
timescales: one is the value derived in the NE2 spectrum and the other
is half the value. 

\begin{figure}
\includegraphics[angle=0,scale=0.65]{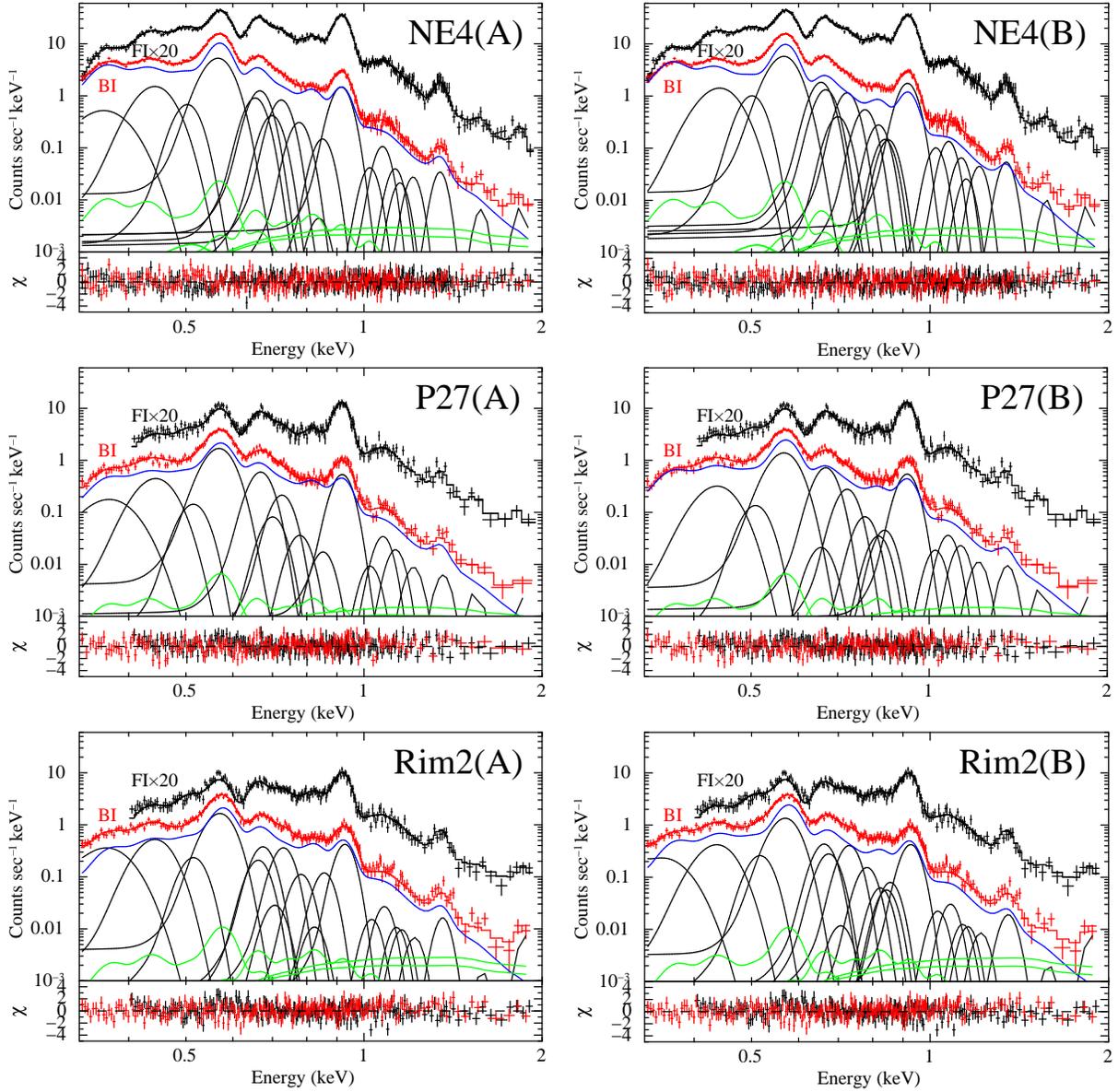}\hspace{1cm}
\caption{Top panels: NE4's outermost rim spectra (black and red for
  the FI and BI data, respectively) along with the best-fit model 
  consisting of an assumed {\tt vpshock} component (blue), 22
  Gaussians (black), and the X-ray background (green).  Contributions 
  of these components are separately shown only for the BI data. 
  The {\tt vpshock} component in the left panel has the
  same spectral parameters derived in the NE2's outermost rim (see,
  Table~\ref{tab:vpshock_param}), whereas that in the right panel has
  half the value for the ionization timescale.  Middle and bottom
  panels: Same as top panels but for P27 (middle) and Rim2 (bottom). 
}
\label{fig:cx}
\end{figure}

In the fitting, we fix all the thermal parameters except for the 
emission measures (EM $=\int n_\mathrm{e}n_\mathrm{H} d\ell$, 
where $\ell$ is the plasma depth).  We introduce 22 Gaussians to  
represent possible CX line emission.  These are K-shell lines from C,
N, O, Ne, Mg, and Si.  For simplicity of the model, we ignore Fe
L-shell lines that might be significant in the X-ray band.  For
example, Fe L-shell lines may affect intensities of the O
Ly($\beta+\gamma+\delta+\epsilon$) around 0.8\,keV, given that the
heliospheric/geocoronal solar-wind charge-exchange (SWCX) emission
shows a hint of an Fe L-shell complex around $\sim$0.8\,keV (Snowden
et al.\ 2004; Fujimoto et al.\ 2007; Smith et al.\ 2007).  Therefore,
the inferred intensities of O Ly($\beta+\gamma+\delta+\epsilon$)
should be considered to be upper limits.  This model gives us
statistically acceptable fits for all the cases examined.
Figure~\ref{fig:cx} shows the best-fit model and the residuals.  In
the figure, left (labeled A) and right (labeled B) columns present
results for the {\tt vpshock} model with the $n_{\mathrm 
e}t$-value of 2.8$\times10^{11}$\,cm$^{-3}$\,sec$^{-1}$ (see, 
Table~\ref{tab:vpshock_param}) and that with half the value,
respectively.  The best-fit EMs (not shown in Table~\ref{tab:cx}) are
obtained to be $\sim$3$\times10^{19}$\,cm$^{-5}$ for NE4, 
$\sim$2.3$\times10^{19}$\,cm$^{-5}$ for P27,
$\sim$1.4$\times10^{19}$\,cm$^{-5}$ for Rim2, respectively. 

\begin{figure}
\includegraphics[angle=0,scale=0.6]{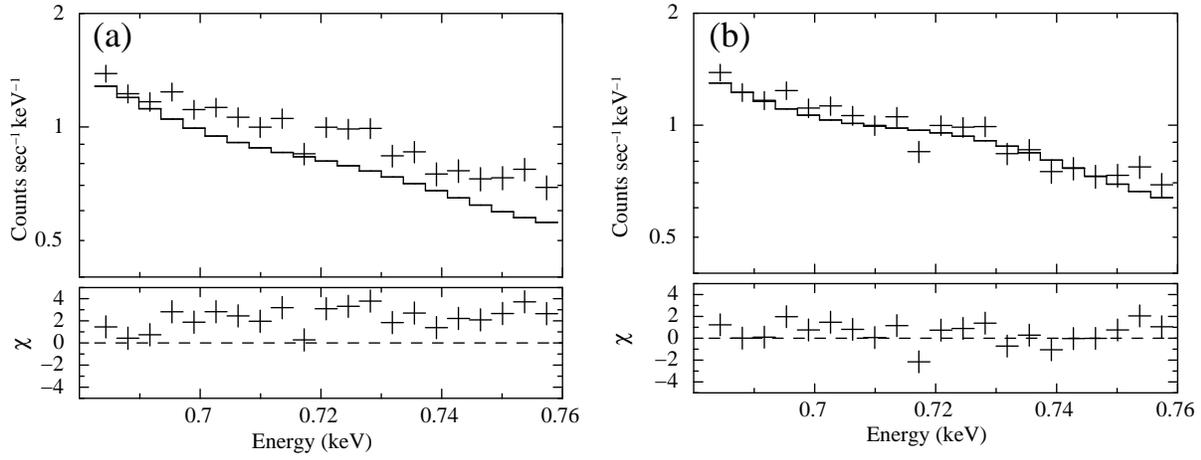}\hspace{1cm}
\caption{(a) The NE4's outermost rim spectrum (only for FI) along with
  the best-fit model consisting of pure thermal emission (see,
  Table~\ref{tab:vpshock_param}).  This is identical to that in
  Fig.~\ref{fig:vpshock} but focusing on a small spectral region of
  0.68--0.76\,keV. (b) Same as Fig.~\ref{fig:zoom} (a) but the model
  includes the possible CX lines.
} 
\label{fig:zoom}
\end{figure}

In order to clarify improvements from the initial fit with the pure 
thermal emission model (Fig.~\ref{fig:vpshock} (a)), we show
the NE4 spectrum with the initial (pure thermal) model in 
Fig.~\ref{fig:zoom} (a) and the new (thermal plus possible CX lines)
model in Fig.~\ref{fig:zoom} (b), focusing on a small spectral region
around the possible CX lines of interest (i.e., 0.68--0.76\,keV).  It
is clear from this figure that the new model fits the data much better
than the initial one.  For quantitative evaluation of the
goodness-of-fits, we perform Kolmogorov-Smirnov (K-S) test for 
the two fit results shown in Fig.~\ref{fig:zoom} (a) and (b).  We find
that the probabilities that the data are drawn from the initial and
new best-fit models are $\sim$15\% and $\sim$80\%, respectively.
Therefore, the new model is much better than the initial one,
although both models cannot be fully ruled out according to this K-S
test.  We also re-fit the data in a wide energy band using a maximum 
likelihood method, and compute the Bayesian Information Criterion
(BIC) defined as $\chi^2+k$ln($n$), where $k$ is the number of free
parameters and $n$ is the number of spectral bins.  The BIC-value is
lower for the new model than the initial one, which also shows the
validity for introducing possible CX lines.

Table~\ref{tab:cx} summarizes the best-fit Gaussian parameters, from
which the intensity ratios of O He($\gamma+\delta+\epsilon$) / O
He$\alpha$ are calculated to be $\sim$0.097--0.110 in NE4,
$\sim$0.057--0.061 in P27, and $\sim$0.066--0.090 in Rim2.  These are 
much higher than that expected for the thermal emission from the
Cygnus Loop ($\sim$0.02 is expected at $kT_{\mathrm e}\sim0.2$\,keV),
but are also larger than those expected in CX emission (the ratio is
likely to be $\sim$0.04; Beiersdorfer et al.\ 2003; Wargelin et al.\
2005).  These inconsistencies imply that our modeling is not perfect
and requires further sophistication which is beyond the scope of this
paper.  Here, it may be worth noting that the ratios derived in our
primitive modeling are closer to CX emission than to thermal emission,
lending support to the idea that CX emission is present. 

Using the line intensities listed in Table~\ref{tab:cx}, we next
estimate ion number ratios in the plasma, as performed for SWCX
emission (e.g., Kharchenko et al.\ 2003; Krasnopolsky 2004; Snowden et
al.\ 2004).  We limit our estimation to only O {\small VIII}, O
{\small IX}, and Ne {\small X}, whose line intensities (K-shell lines
from O {\small VII}, O {\small VIII}, and Ne {\small IX} after CX
reactions) are relatively well constrained.  From Table~\ref{tab:cx},
the intensity ratios of O {\small VIII} lines / O {\small VII}
lines and Ne {\small IX} lines / O {\small VII} lines are calculated
to be $\sim$0.095--0.184 and $\sim$0.079--0.086 in NE4,
$\sim$0.006--0.026 and $\sim$0.059--0.069 in P27, and
$\sim$0.052--0.156 and $\sim$0.040--0.053 in Rim2, respectively.  
Above, we summed K-shell fluxes from all transitions in the same
species, since the total number of K-shell photons reflects the number
of CX reactions.  To convert the flux ratios to ion number ratios, we
use SWCX cross sections for  hydrogen listed in Wargelin et al.\
(2004). The cross section ratios of O {\small IX} / O {\small VIII}
and Ne {\small X} / O {\small VIII} are computed to be 1.66 and 1.53,
respectively.  Dividing the flux ratios by the cross section ratios,
the ion number ratios for O {\small IX} / O {\small VIII} and Ne
{\small X} / O {\small VIII} are derived to be $\sim$0.057--0.111 and
$\sim$0.052--0.056 in NE4, $\sim$0.004--0.016 and $\sim$0.039--0.045
in P27, and $\sim$0.031--0.094 and $\sim$0.026--0.035 in Rim2,
respectively.

We compare these ion number ratios with those expected in the thermal
plasma of the Cygnus Loop's rim, i.e., $kT_{\mathrm e}$=0.2\,keV.
We investigate three ionization timescales: 
$n_{\mathrm e}t$ = (0.5, 1, 2)$\times$10$^{11}$\,cm$^{-3}$\,sec.
As described in the appendix, the ion number ratios for O {\small IX} /
O {\small VIII} and Ne {\small X} / O {\small VIII} are derived to be 
$<$0.33 and 0.01--0.02, respectively (here, we adopt the Ne/O
abundance ratio of 2 times the solar value which is typically reported
in the Cygnus Loop, e.g., Tsunemi et al.\ 2007).  
Although the error range is large for the former ratio, it is
consistent with our estimates ranging from $\sim$0.004 to $\sim$0.111
using SWCX cross sections.  On the other hand, the latter ratio is
significantly lower than our measurement by a factor of
$\sim$1.3--5.6.  This discrepancy suggests either overestimates of
CX lines from He-like Ne, underestimates of CX lines from H-like O, or
both.  The overall intensity of O {\small VIII} lines is also
somewhat uncertain because of the difficulty in separating O 
Ly$\alpha$ at 0.654\,keV and O He$\beta$ at 0.666\,keV as well as the 
possible contribution of Fe L-shell lines around 0.8\,keV.
Based on the best-fit parameters in Tables~\ref{tab:bremss_gauss}
(total emission) and \ref{tab:cx} (CX emission), the fractions of CX
emission in total emission are calculated to be $\sim$40\% for O
He$\alpha$ and $\sim$50\% for Ne He$\alpha$.  A higher fraction of CX
emission in Ne He$\alpha$ than in O He$\alpha$ is difficult to
understand, since the fraction of H-like ions (which yield He$\alpha$
CX emission) is higher for O than for Ne; at the electron temperature
of $\sim$0.2\,keV in thermal equilibrium, a little less than half of O
is He-like, while less than 10\% of Ne is H-like and 
the rest is He-like.  Taking account of NEI effects would make things
worse.  Therefore, Ne He$\alpha$ CX lines are likely to be
overestimated by our primitive CX modeling.  This problem could be
caused by inaccurate modeling of the thermal emission, because the
line ratios strongly depend on the plasma condition which is difficult
to infer in the CX-contaminated spectra.  In addition to the
uncertainty in the spectral modeling, the CX cross sections might be
different from those we employed.  Given these complexities, the
derivation of accurate ion number ratios from CX emission is left for
future work. At the very least, we can argue here that the ion
fractions inferred from both CX emission and the thermal emission
model are consistent within an order of magnitude.

%The overall intensity of O {\small VIII}
%lines is also somewhat uncertain because of the difficulty in
%separating O Ly$\alpha$ at 0.654\,keV and O He$\beta$ at 0.666\,keV as
%well as the possible contribution of Fe L-shell lines around 0.8\,keV.

\subsubsection{CX Emission from SNRs}

Although it is challenging to distinguish between CX emission and
thermal emission in our data, we have constructed a spectral model 
for the CX emission, assuming that spectral shape of the
thermal emission is similar with azimuthal angle.  We here discuss the
inferred CX-to-thermal flux ratios (e.g., O He$\alpha\sim$40\%) in
terms of theoretical expectations.

Wise \& Sarazin (1989) were the first to perform detailed calculations
of CX-induced X-ray emission in a SNR.  They found that CX emission 
generally contributes only 10$^{-3}$ to 10$^{-5}$ of the
collisionally excited lines in the entire SNR, even when the ambient gas
is completely neutral.  For comparison, we estimate the total
CX-to-thermal flux ratio in the Cygnus Loop.  We assume that
the total CX flux is 10 times larger than that in Table~\ref{tab:cx}
for NE4, since the number of FOV where we see CX emission is about ten 
(see, Fig.~\ref{fig:map} (a)).  The total thermal flux from the
remnant is likewise inferred from the integrated thermal flux in all
the spectral extraction regions in Fig.~\ref{fig:image}.  The
CX-to-thermal ratio in the entire SNR is then inferred to be on the
order of 10$^{-2}$, which is significantly higher than that in the
model of Wise \& Sarazin (1989).  This discrepancy requires either a
reduction of the CX contribution in our spectral modeling in
Section~4.3.1, improvements of the theoretical calculation, or perhaps
both.  In fact, physical parameters such as the shock velocity
used in the calculation by Wise \& Sarazin (1989) are different from 
those in the Cygnus Loop.

More recently, Lallement (2004b) investigated spatial variations of 
CX emission in a SNR.  According to Fig.~3 in Lallement (2004b), CX
emission can comprise a few 10\% of the total X-ray (0.1--0.5\,keV)
emission within the radius of $0.98R_\mathrm{shock} < R <
R_\mathrm{shock}$, which corresponds to the outermost rim region in our
analysis.  This theoretical prediction agrees reasonably well with our
measurement of O He$\alpha\sim$40\%.  Although Lallement's calculation
is specifically intended for the LMC SNR DEM~L71, we believe that it
is appropriate for a qualitative comparison with the Cygnus Loop,
since the basic parameters for both the DEM~L71 SNR and the Cygnus
Loop are similar.

Lallement (2004b) also presented the expected EM of thermal emission
at the equivalent radius, i.e., the radius where the flux of CX
emission becomes equal to that of the thermal emission.  The expected
EM is expressed as 
2$\times10^{17}$\,$\epsilon$\,cos$^{-1}\theta$\,$\alpha$\,$\chi$(T,a)$^{-1}$\,$n_\mathrm{c}$\,V$_{100}$\,cm$^{-5}$, 
where $\epsilon$ is the ratio between the CX probability and the
collisional ionization probability, $\theta$ is an incidence angle
(see, Fig.~2 in Lallement 2004b), $\alpha$ accounts for the hot ion
content, i.e., the temperature and metallicity, $\chi$(T,a) is the
thermal emissivity ratio for a given temperature T and the metal
abundance a relative to a temperature 1$\times$10$^{6}$\,K and the
solar abundance, $n_\mathrm{c}$ is the ambient density, and V$_{100}$
is the relative velocity between neutrals and the hot gas in units of 
100\,km\,sec$^{-1}$.  Following the discussion in Lallement (2004b), 
we employ $\epsilon = 1$, $\alpha = 1$, and $\chi$(T,a)$ = 1$. 
The equivalent radius of $\sim0.98R_{\rm shock}$, which is the inner edge
of the outermost region in Fig.~\ref{fig:image} gives us
cos$^{-1}\theta$ of $\sim$5.  The ambient density is taken to be unity
from measured values ranging from 0.3 to 4\,cm$^{-3}$ near NE4
(Raymond et al.\ 2003; Leahy 2003).  The relative velocity can be
considered as the shock velocity of $\sim$300\,km\,sec$^{-1}$
(Ghavamian et al.\ 2001; Salvesen et al.\  2009).  Substituting these
quantities in the equation above, we derive the expected EM of
$\sim5\times10^{18}$\,cm$^{-5}$.  Since the EM observed in the
outermost rim in NE4 is estimated to be
$\sim$1.4--3\,$\times$10$^{19}$\,cm$^{-5}$ (see, Section~4.2.1), the  
expected and observed EMs are consistent within a factor of
$\sim$2--6.  Given the considerable uncertainties in the spectral
modeling as well as the ambient density, we believe that the EMs are
in reasonable agreement.

We have found general agreements between our observational results and
the theoretical predictions by Lallement (2004b).  Nonetheless,
further detailed calculation of the CX emission in a SNR is
encouraged, since Lallement's calculation is somewhat simplified in that
it does not consider the ionization structure of the post-shock region,
and thus would lead to an overestimate of the CX contribution.
Likewise, more sophisticated spectral modeling of CX emission is also
desirable for future work.

\subsection{Need for High-Resolution Spectra with the {\it Astro-H}
  SXS} 

While we have proposed CX emission for the origin of the ``enhanced'' 
abundance in the rim regions of the Cygnus Loop and have presented
supporting observational evidence, solid evidence for the presence of
CX emission is still missing.  Therefore, at this point, we cannot
fully exclude alternative explanations.  For example, there could
really be a bimodal distribution of abundances in the rim of the
Cygnus Loop, in spite of the difficulties described in Section~1.

The {\it Astro-H} SXS will hopefully provide definitive evidence for 
CX emission by resolving the O He$\alpha$ triplets, i.e., resonance, 
intercombination, and forbidden lines.  In thermal emission, the ratio 
of forbidden to resonance line strengths is less than unity.  In contrast, 
for CX emission the ratio is expected to be about 10
(Kharchenko et al.\ 2003).  Laboratory measurements of CX reactions do
show stronger forbidden lines than resonance lines (Beiersdorfer et
al.\ 2003), although the forbidden-to-resonance ratio is not as large
as 10 (which might be related to the fact that the authors used CO$_2$
rather than H as electron-donor species).  Thus, we expect that
forbidden lines will be enhanced in the CX-contaminated outermost rim
regions compared with those in the rest of the rim.  So far, the
northern bright portion of the Cygnus Loop is the only location where
the triplets have been resolved, by high spectral resolution {\it
Einstein} Focal Plane Crystal Spectrometer (FPCS) observations (Vedder
et al.\ 1986).  The resonance line there is much stronger than the 
forbidden line, suggesting that CX emission is not significant.
Since the FPCS FOV (3$^{\prime}\times$30$^{\prime}$) is located
well outside (or inside with respect to the center of the Loop) our
{\it Suzaku} FOV shown in Fig~\ref{fig:image}, it is reasonable to
consider that signatures of CX emission if present would be washed out
by thermal emission.  Spatial variations of the line ratio, which
provide the key information about the presence of CX emission, will be
measured with the {\it Astro-H} SXS. 

With the current CCD data, it might be possible to find a
forbidden/resonance line-ratio variation by measuring the energy
separation between O He$\alpha$ and O Ly$\alpha$ (e.g., Lallement
2009).  However, such energy separations are expected to be 
less than a few eV, and thus are too small to detect with
nondispersive CCD spectra.  As we show in
Tables~\ref{tab:bremss_gauss} and \ref{tab:cx}, the fit line center
energies in the {\it Suzaku} spectra are randomly scattered 
within the energy-scale calibration
uncertainties ($\pm$5\,eV: Koyama et al.\ 2007).  
Moreover, O Ly$\alpha$ is significantly contaminated by O
He$\beta$ in part of the Cygnus Loop, and its contribution is highly
variable from location to location.  This makes even more difficult any 
firm measurement of the separation between O He$\alpha$ and O
Ly$\alpha$.  Also, dispersed spectra from slitless X-ray grating spectrometers
are almost useless for the Cygnus 
Loop because of its large extent, 
although the {\it XMM-Newton} RGS
might provide better spectra for some bright knotty features.  Since
the {\it Astro-H} SXS is a nondispersive microcalorimeter, it will be
the first instrument to provide us with high-resolution spectra of
sufficient quality to resolve individual lines of the triplets.  The
high-resolution spectra will also make possible identification of the
individual lines around 0.7\,keV.  Therefore, the {\it Astro-H} SXS is
being eagerly awaited for further studies of the Cygnus Loop.

In addition, the {\it Astro-H} SXS will help us derive much more
accurate absolute metal abundances than those from nondispersive CCD
spectra.  This is because we can directly measure equivalent widths
of lines below 1\,keV with SXS spectra, which is not possible with CCD 
spectra.  

While measuring the absolute abundance with CCD spectra is 
quite difficult, it should be noted that relative metal abundances 
are already well constrained unless the spectra are heavily
contaminated by CX emission.  Our analysis suggests that CX
emission is not significant in the ``normal'' abundance regions. 
The relative abundances obtained there are close to the solar ratio
except for O: the abundance of O is depleted by an additional factor
of 2 relative to other elements (e.g., Miyata et al.\ 1994).  As
Miyata et al.\ (2008) pointed out, the depletion of the O abundance
might be due to resonance-line scattering which is especially
important for O K-shell lines.  Therefore, in order to obtain accurate
absolute and relative abundances in the rim of the Cygnus Loop, we
will need to carefully consider effects of both CX emission and
resonance-line scattering.

\section{Conclusion}

Using {\it Suzaku} and {\it XMM-Newton} data covering nearly the
entire rim of the Cygnus Loop, we found that outermost rims in N--NE,
SE, and NW regions show a strong line feature at $\sim$0.7\,keV.  Its
anomalously strong intensity and spatial localization in the outermost
rim lead us to propose that it originates from CX processes, most
likely between neutrals and H-like O ions.  If so, this is the first
detection of CX-induced X-ray emission associated with the forward
shock of a SNR.  The regions showing the $\sim$0.7\,keV emission
correspond to the ``enhanced'' abundance regions, and we suggest that
the $\sim$0.7\,keV emission and possibly other CX lines could enhance
the apparent metal abundances there when we fit the CX-contaminated 
spectra with pure thermal emission models.  The introduction of the CX
emission potentially resolves the mysterious abundance
``enhancements'' in the outermost rim regions of the Cygnus Loop.  On 
the other hand, the abundance depletion in overall rim regions still
remains an open question.

%Finally, we point out that this work adds the Cygnus Loop to the
%growing list of objects where CX plays a non-negligible role.  CX is
%thought now to have been found in the North Polar Spur, the Carina
%nebula, and the bulge of M31.

\acknowledgments

The authors are grateful to Una Hwang and the anonymous referee for
numerous constructive and insightful comments which significantly
improved the quality of this paper.  S.K.\ is supported by a JSPS
Research Fellowship for Research Abroad, and in part by the NASA grant
under the contract NNG06EO90A.  M.K.\ is supported by JSPS Research 
Fellowship for Young Scientists.

\section{Appendix}
%\appendix

To estimate the ion fractions, we first calculate ion fractions at 
collisional ionization equilibrium using {\tt
  sherpa}\footnote{http://cxc.harvard.edu/ciao/sherpa}.  Second, we 
calculate $n_z$/$n_\mathrm{Z}$$\epsilon_{i}(T)$ (i.e., a product of
ion fraction and emissivity) of some selected lines as a function of
$n_{\mathrm e} t$, using the {\tt NEIline} software 
provided by R.\ Smith (applications of this software can be found in
literature, e.g., Sasaki et al.\ 2004; Katsuda \& Tsunemi 2006).
Above, $n_z(n_{\mathrm e} t)$/$n_\mathrm{Z}$ is the ionization
fraction of the ionization species responsible for the line, and
$\epsilon_{i}(T)$, a function of temperature, is the intrinsic
emissivity of the line.  Then, assuming that the $\epsilon_{i}(T)$ is
constant with $n_{\mathrm e}t$, we roughly estimate ion fractions at
various $n_{\mathrm e} t$-values.  To evaluate systematic
uncertainties associated with this method, we use not only K$\alpha$
lines but also K$\beta$ lines.  This method does not directly give ion
fractions for fully ionized ions, since they do not radiate line
emission by collisional excitation.  We thus assume that fully ionized
ions take up the balance, which is a good approximation in the plasma
condition of interest.

\begin{deluxetable}{lcccc}
\tabletypesize{\scriptsize}
\tablecaption{Summary of observations.}
\tablewidth{0pt}
\tablehead{
\colhead{Obs.\ ID}&\colhead{Obs.\ Date}&\colhead{R.A., Decl.\
  (J2000)}&\colhead{Position Angle}&\colhead{Effective Exposure}}
\startdata
500020010 (NE1)\dotfill & 2005-11-23 & 20$^{h}$56$^{m}$45$^s$.24, 31$^\circ$44$^{\prime}$16$^{\prime\prime}$.8 & 223$^\circ$.0 & 20.4\,ksec\\ 
500021010 (NE2)\dotfill & 2005-11-24 & 20$^{h}$55$^{m}$52$^s$.34, 31$^\circ$57$^{\prime}$15$^{\prime\prime}$.1 & 223$^\circ$.0 & 21.4\,ksec\\ 
500022010 (NE3)\dotfill & 2005-11-29 & 20$^{h}$55$^{m}$01$^s$.99, 32$^\circ$10$^{\prime}$57$^{\prime\prime}$.4 & 222$^\circ$.9 & 21.7\,ksec\\ 
500023010 (NE4)\dotfill & 2005-11-30 & 20$^{h}$54$^{m}$00$^s$.12, 32$^\circ$22$^{\prime}$08$^{\prime\prime}$.4 & 221$^\circ$.2 & 25.3\,ksec\\ 
501020010 (P10)\dotfill & 2007-11-26 & 20$^{h}$46$^{m}$17$^s$.86, 30$^\circ$23$^{\prime}$57$^{\prime\prime}$.1 & 240$^\circ$.0 & 16.8\,ksec\\ 
501035010 (P18)\dotfill & 2006-12-18 & 20$^{h}$48$^{m}$13$^s$.13, 29$^\circ$42$^{\prime}$40$^{\prime\prime}$.0 & 237$^\circ$.5 & 12.0\,ksec\\ 
501036010 (P19)\dotfill & 2006-12-18 & 20$^{h}$47$^{m}$14$^s$.26, 30$^\circ$04$^{\prime}$54$^{\prime\prime}$.5 & 237$^\circ$.5 & 18.6\,ksec\\ 
503057010 (P21)\dotfill & 2008-06-02 & 20$^{h}$52$^{m}$47$^s$.04, 32$^\circ$25$^{\prime}$10$^{\prime\prime}$.9 & 61$^\circ$.9 & 16.2\,ksec\\ 
503058010 (P22)\dotfill & 2008-06-03 & 20$^{h}$51$^{m}$20$^s$.47, 32$^\circ$24$^{\prime}$16$^{\prime\prime}$.9 & 61$^\circ$.4 & 19.3\,ksec\\ 
503059010 (P23)\dotfill & 2008-06-03 & 20$^{h}$49$^{m}$54$^s$.53, 32$^\circ$21$^{\prime}$31$^{\prime\prime}$.3 & 61$^\circ$.9 & 19.5\,ksec\\ 
503060010 (P24)\dotfill & 2008-06-04 & 20$^{h}$48$^{m}$32$^s$.16, 32$^\circ$17$^{\prime}$25$^{\prime\prime}$.8 & 61$^\circ$.4 & 18,5\,ksec\\ 
503061010 (P25)\dotfill & 2008-06-04 & 20$^{h}$47$^{m}$26$^s$.59, 32$^\circ$10$^{\prime}$04$^{\prime\prime}$.1 & 60$^\circ$.9 & 26.0\,ksec\\ 
503062010 (P26)\dotfill & 2008-05-13 & 20$^{h}$56$^{m}$30$^s$.05, 30$^\circ$18$^{\prime}$48$^{\prime\prime}$.6 & 49$^\circ$.8 & 16.9\,ksec\\ 
503063010 (P27)\dotfill & 2008-05-13 & 20$^{h}$55$^{m}$19$^s$.87, 30$^\circ$00$^{\prime}$37$^{\prime\prime}$.4 & 49$^\circ$.6 & 22.8\,ksec\\ 
503064010 (P28)\dotfill & 2008-05-14 & 20$^{h}$53$^{m}$55$^s$.13, 29$^\circ$53$^{\prime}$36$^{\prime\prime}$.2 & 49$^\circ$.1 & 18.2\,ksec\\ 
504005010 (Rim1)\dotfill & 2009-11-17 & 20$^{h}$46$^{m}$34$^s$.10, 31$^\circ$52$^{\prime}$58$^{\prime\prime}$.8 & 60$^\circ$.0 & 40.7\,ksec\\ 
504006010 (Rim2)\dotfill & 2009-11-18 & 20$^{h}$45$^{m}$42$^s$.24, 31$^\circ$35$^{\prime}$40$^{\prime\prime}$.6 & 60$^\circ$.0 & 26.3\,ksec\\ 
504007010 (Rim3)\dotfill & 2009-11-19 & 20$^{h}$45$^{m}$17$^s$.57, 31$^\circ$17$^{\prime}$57$^{\prime\prime}$.5 & 246$^\circ$.4 & 21.6\,ksec\\ 
504008010 (Rim4)\dotfill & 2009-11-20 & 20$^{h}$45$^{m}$52$^s$.27, 31$^\circ$00$^{\prime}$47$^{\prime\prime}$.2 & 246$^\circ$.0 & 12.1\,ksec\\ 
504009010 (Rim5)\dotfill & 2009-11-20 & 20$^{h}$46$^{m}$06$^s$.86, 30$^\circ$40$^{\prime}$52$^{\prime\prime}$.7 & 216$^\circ$.0 & 15.9\,ksec\\ 
504010010 (Rim6)\dotfill & 2009-11-20 & 20$^{h}$57$^{m}$30$^s$.50, 31$^\circ$27$^{\prime}$01$^{\prime\prime}$.1 & 220$^\circ$.0 & 14.3\,ksec\\ 
0018141301 ({\it XMM-Newton})\dotfill & 2002-04-30 & 20$^{h}$57$^{m}$21$^s$.02, 31$^\circ$00$^{\prime}$13$^{\prime\prime}$.3 & 262$^\circ$.4 & 11.2\,ksec\\ 
\enddata
\label{tab:obs}
\end{deluxetable}

\begin{deluxetable}{lccc}
\tabletypesize{\scriptsize}
\tablecaption{Spectral-fit parameters in the outermost regions of NE4
  and NE2, using the absorbed {\tt vpshock} model.}
\tablewidth{0pt}
\tablehead{
\colhead{Parameter}&\colhead{NE4}&\colhead{NE4 (ignoring
  0.68--0.76\,keV)}&\colhead{NE2}
}
\startdata
$N_\mathrm{H}$[$\times10^{22}$cm$^{-2}$]\dotfill & 0.03 (fixed) & 0.03
(fixed) & 0.03 (fixed)  \\
$kT_\mathrm{e}$[keV]\dotfill & 0.34$^{+0.01}_{-0.03}$ & 0.22$\pm$0.01
& 0.21$\pm$0.01 \\
log$(n_\mathrm{e}t_\mathrm{upper}/\mathrm{cm}^{-3}\,\mathrm{sec})$\dotfill 
&10.77$^{+0.11}_{-0.03}$&11.41$^{+0.06}_{-0.03}$ &
11.44$^{+0.10}_{-0.08}$ \\  
C\dotfill  & 0.85$^{+0.11}_{-0.22}$&0.20$^{+0.02}_{-0.04}$ &
0.17$^{+0.04}_{-0.03}$ \\ 
N\dotfill  & 0.88$^{+0.11}_{-0.25}$ & 0.15$^{+0.02}_{-0.03}$&
0.07$\pm$0.02 \\ 
O\dotfill  & 0.43$^{+0.05}_{-0.11}$ &0.12$^{+0.01}_{-0.02}$
&0.08$\pm$0.01 \\ 
Ne\dotfill  & 0.75$^{+0.09}_{-0.16}$ &0.29$^{+0.02}_{-0.03}$ &
0.15$\pm$0.02 \\ 
Mg\dotfill  & 0.29$^{+0.04}_{-0.05}$&0.12$\pm$0.02
&0.10$^{+0.03}_{-0.04}$ \\ 
Si(=S)\dotfill  & 0.22$^{+0.11}_{-0.12}$&$<$0.11 & $<$0.10 \\ 
Fe(=Ni)\dotfill  & 0.38$^{+0.05}_{-0.09}$&0.11$\pm$0.01 &
0.14$\pm$0.02\\ 
EM$^{a}$[$\times10^{19}$ cm$^{-5}$]\dotfill & 0.36$^{+0.20}_{-0.04}$ &
2.9$^{+0.5}_{-0.2}$ & 1.5$\pm$0.3 \\ 
Energy shifts[eV]\dotfill & $+$0.4 (FI), $-$1.7 (BI) & $+$0.4 (FI),
$-$1.7 (BI) &$-$2.9 (FI), $+$0.2 (BI)\\  
\hline
$\chi^2$/d.o.f. \dotfill & 1219/551 & 802/505 & 466/410\\
\enddata
\tablecomments{$^{a}$EM denotes the emission measure ($=\int
  n_\mathrm{e}n_\mathrm{H} d\ell$).  The values of 
  abundances are multiples of the solar values (Anders \& Grevesse
  1989).  The errors represent 90\% confidence ranges.} 

\label{tab:vpshock_param}
\end{deluxetable}

\begin{deluxetable}{lcccccc}
\tabletypesize{\scriptsize}
\tablecaption{Gaussian parameters in the NE4, NE2, P27, P19, Rim2,
  and P24's outermost rims, using an absorbed phenomenological model 
  consisting of an underlying bremsstrahlung plus 18 Gaussians.} 
\tablewidth{0pt}
\tablehead{
\colhead{Line}&\colhead{NE4$^{a}$}&\colhead{NE2}&\colhead{P27$^{a}$}&\colhead{P19}&\colhead{Rim2$^{a}$}&\colhead{P24}} 
\startdata
C Ly$\alpha$ \\
~~Center (eV)\dotfill & 357(357)$^{+2}_{-3}$ & 351(354)$\pm5$ & 349(351)$\pm8$ & 349(345)$\pm6$ & 349(356)$\pm3$ & 347(348)$\pm6$ \\
~~Width (eV)\dotfill & 28$\pm$2 & 30$\pm$3 & 32$^{+3}_{-5}$ & 32$^{+2}_{-4}$ & 30$\pm$2 & 29$^{+3}_{-4}$ \\
~~Normalization$^{b}$\dotfill & 7606$^{+568}_{-405}$ & 3215$^{+309}_{-324}$ & 7998$^{+500}_{-1363}$ & 9550$^{+482}_{-1289}$ & 7074$^{+396}_{-504}$ & 5421$^{+456}_{-778}$ \\
\hline
C Ly$\beta$ + N He$\alpha$ \\
~~Center (eV)\dotfill & 438(438)$\pm$1 & 437(441)$\pm$2 & 435(438)$^{+2}_{-4}$ & 432(429)$\pm$2 & 435(442)$\pm$3 & 433(434)$^{+3}_{-5}$ \\
~~Width (eV)\dotfill & 14$^{+1}_{-2}$ & 14$\pm$3 & 14$^{+4}_{-2}$ & 14$^{+2}_{-3}$ & 11$^{+3}_{-7}$ & 9$^{+6}_{-10}$ \\
~~Normalization$^{b}$\dotfill & 2023$^{+51}_{-63}$ & 591$^{+32}_{-18}$ & 1779$^{+128}_{-116}$ & 2239$^{+126}_{-127}$ & 958$^{+83}_{-90}$ & 822$^{+95}_{-75}$ \\
\hline
N Ly$\alpha$ + N He$\beta$ \\
~~Center (eV)\dotfill & 499(499)$^{+1}_{-2}$ & 497(501)$^{+0}_{-4}$ & 499(501)$^{+2}_{-4}$ & 499(496)$^{+1}_{-2}$ & 499(506)$^{+2}_{-3}$ & 495(496)$^{+3}_{-2}$ \\
~~Width (eV)\dotfill & 0$^{c}$ & 0$^{c}$ & 0$^{c}$ & 0$^{c}$ & 0$^{c}$ & 0$^{c}$ \\
~~Normalization$^{b}$\dotfill & 730$\pm$20 & 177$\pm$12 & 481$\pm$38 & 686$\pm$39 & 281$\pm$23 & 257$\pm$25 \\
\hline
O He$\alpha$ \\
~~Center (eV)\dotfill & 565(565)$\pm$0 & 564(567)$\pm$1 & 564(567)$\pm$1 & 565(561)$\pm$1 & 564(571)$\pm$1 & 563(564)$\pm$1 \\
~~Width (eV)\dotfill & 12$\pm$0 & 12$\pm$1 & 13$\pm$1 & 11$^{+1}_{-2}$ & 12$\pm$1 & 12$^{+1}_{-2}$ \\
~~Normalization$^{b}$\dotfill & 5547$^{+53}_{-55}$ & 1659$\pm$34 & 4763$^{+98}_{-106}$ & 4559$^{+97}_{-105}$ & 2804$^{+62}_{-59}$ & 2243$^{+69}_{-70}$ \\
\hline
O Ly$\alpha$ \\
~~Center (eV)\dotfill & 654(653)$^{c}$ & 651(655)$^{c}$ & 653(655)$^{c}$ & 653(649)$^{c}$ & 653(660)$^{c}$ & 651(652)$^{c}$ \\
~~Width (eV)\dotfill & 0$^{c}$ & 0$^{c}$ & 0$^{c}$ & 0$^{c}$ & 0$^{c}$ & 0$^{c}$ \\
~~Normalization$^{b}$\dotfill & 602$\pm$17 & 208$\pm$10 & 258$\pm$28 & 607$\pm$32 & 259$\pm$16 & 178$\pm$18 \\
\hline
O He$\beta$ \\
~~Center (eV)\dotfill & 666(665)$^{c}$ & 663(667)$^{c}$ & 665(667)$^{c}$ & 665(661)$^{c}$ & 665(672)$^{c}$ & 663(664)$^{c}$ \\
~~Width (eV)\dotfill & 0$^{c}$ & 0$^{c}$ & 0$^{c}$ & 0$^{c}$ & 0$^{c}$ & 0$^{c}$ \\
~~Normalization$^{b}$\dotfill & 367$\pm$16 & 82$\pm$9 & 444$\pm$26 & 437$\pm$29 & 130$\pm$14 & 194$\pm$16 \\
\hline
O He$\gamma+$etc. \\
~~Center (eV)\dotfill & 704(704)$\pm$2 & 708(711)$^{+7}_{-8}$ & 703(706)$\pm$5 & 711(708)$\pm$6 & 705(712)$^{+4}_{-5}$ & 718(719)$^{+8}_{-9}$ \\
~~Width (eV)\dotfill & 22$\pm$2 & 20$^{+11}_{-10}$ & 24$^{+5}_{-6}$ & 23$\pm$7 & 23$^{+5}_{-6}$ & 0$^{+24}_{-0}$ \\
~~Normalization$^{a}$\dotfill & 341$^{+17}_{-16}$ & 48$^{+10}_{-9}$ & 239$^{+28}_{-27}$ & 206$^{+31}_{-29}$ & 145$^{+16}_{-15}$ & 47$^{+14}_{-9}$ \\
\hline
Fe L (3d$\to$2p) \\
~~Center (eV)\dotfill & 726(725)$^{c}$ & 723(727)$^{c}$ & 725(727)$^{c}$ & 725(721)$^{c}$ & 725(732)$^{c}$ & 723(724)$^{c}$ \\
~~Width (eV)\dotfill & 0$^{c}$ & 0$^{c}$ & 0$^{c}$ & 0$^{c}$ & 0$^{c}$ & 0$^{c}$ \\
~~Normalization$^{b}$\dotfill & \multicolumn{6}{c}{Linked to Fe L (3d$\to$2p)} \\
\hline
O Ly$\beta$ \\
~~Center (eV)\dotfill & 775(774)$^{c}$ & 772(776)$^{c}$ & 774(776)$^{c}$ & 774(770)$^{c}$ & 774(781)$^{c}$ & 772(773)$^{c}$ \\
~~Width (eV)\dotfill & 0$^{c}$ & 0$^{c}$ & 0$^{c}$ & 0$^{c}$ & 0$^{c}$ & 0$^{c}$ \\
~~Normalization$^{b}$\dotfill & 99$\pm$6 & 35$\pm$4 & 54$\pm$8 & 134$\pm$12 & 38$\pm$5 & 42$\pm$6 \\
\hline
Fe L (3s$\to$2p) \\
~~Center (eV)\dotfill & 822(822)$\pm$1 & 819(823)$^{+1}_{-2}$ & 819(821)$^{+2}_{-3}$ & 822(818)$^{+2}_{-3}$ & 822(829)$^{+3}_{-2}$ & 819(820)$\pm$2 \\
~~Width (eV)\dotfill & 0$^{c}$ & 0$^{c}$ & 0$^{c}$ & 0$^{c}$ & 0$^{c}$ & 0$^{c}$ \\
~~Normalization$^{b}$\dotfill & 138$\pm$4 & 57$\pm$3 & 93$\pm$6 & 247$\pm$9 & 67$\pm$4 & 64$\pm$4 \\
\hline
Ne He$\alpha$ \\
~~Center (eV)\dotfill & 912(912)$\pm$1 & 909(912)$^{+1}_{-2}$ & 913(915)$\pm$1 & 912(908)$\pm$1 & 915(921)$\pm$1 & 912(913)$\pm$2 \\
~~Width (eV)\dotfill & 15$\pm$1 & 19$\pm$2 & 13$\pm$2 & 15$\pm$2 & 14$\pm$2 & 13$^{+3}_{-4}$ \\
~~Normalization$^{b}$\dotfill & 364$^{+5}_{-6}$ & 89$^{+4}_{-3}$ & 282$\pm$8 & 337$\pm$10 & 140$\pm$4 & 111$\pm$5 \\
\hline
\\
\\
Ne Ly$\alpha$ \\
~~Center (eV)\dotfill & 1022(1021)$^{c}$ & 1019(1023)$^{c}$ & 1021(1023)$^{c}$ & 1021(1017)$^{c}$ & 1021(1028)$^{c}$ & 1019(1020)$^{c}$ \\
~~Width (eV)\dotfill & 0$^{c}$ & 0$^{c}$ & 0$^{c}$ & 0$^{c}$ & 0$^{c}$ & 0$^{c}$ \\
~~Normalization$^{b}$\dotfill & 16$\pm$2 & 7$\pm$1 & 12$\pm$2 & 26$\pm$3 & 7$\pm$1 & 10$\pm$2 \\
\hline
Ne He$\beta$ \\
~~Center (eV)\dotfill & 1074(1073)$^{c}$ & 1071(1075)$^{c}$ & 1073(1075)$^{c}$ & 1073(1069)$^{c}$ & 1073(1080)$^{c}$ & 1071(1072)$^{c}$ \\
~~Width (eV)\dotfill & 0$^{c}$ & 0$^{c}$ & 0$^{c}$ & 0$^{c}$ & 0$^{c}$ & 0$^{c}$ \\
~~Normalization$^{b}$\dotfill & 19$\pm$1 & 5$\pm$1 & 15$\pm$2 & 16$\pm$3 & 7$\pm$1 & 6$\pm$1 \\
\hline
Ne He$\gamma$ \\
~~Center (eV)\dotfill & 1127(1126)$^{c}$ & 1124(1128)$^{c}$ & 1126(1128)$^{c}$ & 1126(1122)$^{c}$ & 1126(1133)$^{c}$ & 1124(1125)$^{c}$ \\
~~Width (eV)\dotfill & 0$^{c}$ & 0$^{c}$ & 0$^{c}$ & 0$^{c}$ & 0$^{c}$ & 0$^{c}$ \\
~~Normalization$^{b}$\dotfill & 12$\pm$1 & 2$\pm$1 & 8$\pm$2 & 17$\pm$3 & 5$^{+1}_{-2}$ & 5$\pm$1 \\
\hline
Ne Ly$\beta$ \\
~~Center (eV)\dotfill & 1210(1209)$^{c}$ & 1207(1211)$^{c}$ & 1209(1211)$^{c}$ & 1209(1205)$^{c}$ & 1209(1216)$^{c}$ & 1207(1208)$^{c}$ \\
~~Width (eV)\dotfill & 0$^{c}$ & 0$^{c}$ & 0$^{c}$ & 0$^{c}$ & 0$^{c}$ & 0$^{c}$ \\
~~Normalization$^{b}$\dotfill & 5$\pm$1 & 3$\pm$1 & 7$\pm$2 & 5$^{+3}_{-2}$ & 4$\pm$1 & 4$\pm$1 \\
\hline
Mg He$\alpha$ \\
~~Center (eV)\dotfill & 1341(1341)$\pm4$ & 1346(1350)$\pm5$ & 1354(1356)$\pm11$ & 1343(1339)$\pm5$ & 1350(1357)$\pm7$ & 1346(1347)$\pm8$ \\
~~Width (eV)\dotfill & 12 ($<$19) & 0 ($<$31) & 36$^{+35}_{-22}$ & 6
($<$22) & 27$^{+10}_{-13}$ & 39$^{+39}_{-13}$ \\ 
~~Normalization$^{b}$\dotfill & 8.2$\pm$1.1 & 2.9$\pm$1.0 & 6.2$^{+1.4}_{-1.5}$ & 9.9$\pm$1.8 & 4.8$^{+1.0}_{-1.1}$ & 6.6$\pm$1.3 \\
\hline
Mg He$\beta$ \\
~~Center (eV)\dotfill & 1571(1571)$\pm17$ & 1511(1514)$\pm38$ &
1522(1525)$\pm381$ & 1578(1574)$\pm395$ & 1574(1581)$\pm394$ &
1576(1577)$\pm394$ \\ 
~~Width (eV)\dotfill & 0$^{c}$ & 0$^{c}$ & 0$^{c}$ & 0$^{c}$ & 0$^{c}$ & 0$^{c}$ \\
~~Normalization$^{b}$\dotfill & 1.2$\pm$0.8 & 0.5 ($<$1.2) &
1.52$^{+1.0}_{-1.2}$ & 1.2 ($<$2.4) & 0.8 ($<$1.6) & 0.9 ($<$1.8) \\ 
\hline
Si He$\alpha$ \\
~~Center (eV)\dotfill & 1833(1833)$^{+25}_{-26}$ & 1886(1889)$\pm22$ &
1839(1841)$\pm57$ & 1865(1861)$\pm18$ & 1872(1879)$\pm37$ &
1806(1808)$\pm43$ \\   
~~Width (eV)\dotfill & 0$^{c}$ & 0$^{c}$ & 0$^{c}$ & 0$^{c}$ & 0$^{c}$ & 0$^{c}$ \\
~~Normalization$^{b}$\dotfill & 1.3$\pm$0.9 & 0.8 ($<$1.7) & 1.4 ($<$2.4) &
2.8$^{+2.5}_{-1.3}$ & 1.5$^{+1.0}_{-1.1}$ & 1.1$\pm$0.9 \\ 
\enddata
\tablecomments{$^{a}$''Enhanced''-abundance regions. $^{b}$In units of
  photons\,cm$^{-2}$\,s$^{-1}$\,sr$^{-1}$.  $^b$Fixed values.  Line
  center energies include effects of energy shifts determined by the
  fittings.  Values in parentheses are derived from the BI CCD. The
  errors (90\% confidence level) are estimated with the other Gaussian 
  parameters and the bremsstrahlung parameters fixed at the best-fit
  values.  These errors do not include systematic uncertainties, such
  as $\pm$5\,eV for the line center energies (Koyama et al.\ 2007).  
}
\label{tab:bremss_gauss}
\end{deluxetable}

\begin{deluxetable}{lcc|cc|cc}
\tabletypesize{\scriptsize}
\tablecaption{Gaussian parameters for the possible CX emission in the
  NE4, P27, and Rim2's outermost rims, using an absorbed combination
  model consisting of the {\tt vpshock} plus 22 Gaussians.}
\tablewidth{0pt}
\tablehead{
\colhead{Line}&\multicolumn{2}{c}{NE4}&\multicolumn{2}{c}{P27}&\multicolumn{2}{c}{Rim2}}
\startdata
& A & B & A & B & A & B\\
C Ly$\alpha$ \\
~~Center (eV)\dotfill & 343(343)$\pm13$ & --- & 345(347)$\pm17$ & --- & 330(337)$^{+11}_{-6}$ & 313(320)$^{+17}_{-9}$ \\
~~Width (eV)\dotfill & 30$^{+29}_{-10}$ & --- & 30$^{+29}_{-11}$ & --- & 30$^{+29}_{-3}$ & 29$^{+29}_{-4}$ \\
~~Normalization$^{a}$\dotfill & 1121$^{+477}_{-438}$ & --- & 2728$^{+1675}_{-1019}$ & --- & 6346$^{+642}_{-1539}$ & 7485$^{+1216}_{-2916}$ \\
\hline
C Ly$\beta$ + N He$\alpha$ \\
~~Center (eV)\dotfill & 436(436)$^{+2}_{-4}$ & 433(433)$\pm$3 & 435(437)$^{+3}_{-9}$ & 426(428)$^{+8}_{-7}$ & 428(435)$^{+4}_{-6}$ & 419(426)$^{+8}_{-6}$ \\
~~Width (eV)\dotfill & 20$^{+5}_{-3}$ & 24$^{+4}_{-3}$ & 20$^{+8}_{-5}$ & 22$^{+8}_{-9}$ & 18$^{+6}_{-5}$ & 21$^{+7}_{-8}$ \\
~~Normalization$^{a}$\dotfill & 869$^{+49}_{-50}$ & 922$^{+104}_{-52}$ & 1082$^{+202}_{-105}$ & 923$^{+188}_{-192}$ & 851$^{+148}_{-75}$ & 818$^{+153}_{-142}$ \\
\hline
N Ly$\alpha$ + N He$\beta$ \\
~~Center (eV)\dotfill & 500(500)$\pm$3 & 499(499)$\pm$2 & 510(512)$^{+-511}_{-9}$ & 503(506)$^{+11}_{-12}$ & 505(512)$\pm$7 & 507(514)$^{+5}_{-7}$ \\
~~Width (eV)\dotfill & 0$^{b}$ & 0$^{b}$ & 0$^{b}$ & 0$^{b}$ & 0$^{b}$ & 0$^{b}$ \\
~~Normalization$^{a}$\dotfill & 160$\pm$21 & 256$^{+21}_{-23}$ & 113$\pm$37 & 114$^{+39}_{-38}$ & 110$^{+23}_{-24}$ & 125$^{+25}_{-24}$ \\
\hline
O He$\alpha$ \\
~~Center (eV)\dotfill & 563(563)$\pm$1 & 563(563)$\pm$1 & 562(565)$\pm$1 & 562(564)$\pm$1 & 560(567)$\pm$1 & 559(566)$\pm$2 \\
~~Width (eV)\dotfill & 0 & 0 & 0 & 0 & 0 & 0 \\
~~Normalization$^{a}$\dotfill & 2058$^{+65}_{-64}$ & 2307$^{+67}_{-66}$ & 2262$^{+119}_{-121}$ & 1926$\pm$125 & 1281$\pm$71 & 1054$\pm$75 \\
\hline
O Ly$\alpha$ \\
~~Center (eV)\dotfill & 654(653)$^{b}$ & 654(653)$^{b}$ & 653(655)$^{b}$ & 653(655)$^{b}$ & 653(660)$^{b}$ & 653(660)$^{b}$ \\
~~Width (eV)\dotfill & 0$^{b}$ & 0$^{b}$ & 0$^{b}$ & 0$^{b}$ & 0$^{b}$ & 0$^{b}$ \\
~~Normalization$^{a}$\dotfill & 180$^{+58}_{-73}$ & 369$^{+63}_{-65}$ &
0 ($<$76) & 10 ($<$138) & 45 ($<$121) & 134$^{+73}_{-49}$ \\
\hline
O He$\beta$ \\
~~Center (eV)\dotfill & 666(665)$^{b}$ & 666(665)$^{b}$ & 665(667)$^{b}$ & 665(667)$^{b}$ & 665(672)$^{b}$ & 665(672)$^{b}$ \\
~~Width (eV)\dotfill & 0$^{b}$ & 0$^{b}$ & 0$^{b}$ & 0$^{b}$ & 0$^{b}$ & 0$^{b}$ \\
~~Normalization$^{a}$\dotfill & 226$^{+90}_{-67}$ & 246$^{+69}_{-74}$ &
324$^{+45}_{-88}$ & 396$^{+38}_{-146}$ & 125$^{+45}_{-88}$ & 86
($<$133) \\ 
\hline
O He$\gamma$ \\
~~Center (eV)\dotfill & 698(697)$^{b}$ & 698(697)$^{b}$ & 697(699)$^{b}$ & 697(699)$^{b}$ & 697(704)$^{b}$ & 697(704)$^{b}$ \\
~~Width (eV)\dotfill & 0$^{b}$ & 0$^{b}$ & 0$^{b}$ & 0$^{b}$ & 0$^{b}$ & 0$^{b}$ \\
~~Normalization$^{a}$\dotfill & 70 ($<$102) & 70$^{+35}_{-16}$ & 39 ($<$83) & 0
($<$63) & 0 ($<$41) & 0 ($<$42) \\
\hline
O He$\delta$ \\
~~Center (eV)\dotfill & 713(712)$^{b}$ & 713(712)$^{b}$ & 712(714)$^{b}$ & 712(714)$^{b}$ & 712(719)$^{b}$ & 712(719)$^{b}$ \\
~~Width (eV)\dotfill & 0$^{b}$ & 0$^{b}$ & 0$^{b}$ & 0$^{b}$ & 0$^{b}$ & 0$^{b}$ \\
~~Normalization$^{a}$\dotfill & 0 ($<$110) & 0 ($<$67) & 0 ($<$109) & 0 ($<$87)
& 0 ($<$59) & 0 ($<$58) \\
\hline
O He$\epsilon$ \\
~~Center (eV)\dotfill & 723(722)$^{b}$ & 723(722)$^{b}$ & 722(724)$^{b}$ & 722(724)$^{b}$ & 722(729)$^{b}$ & 722(729)$^{b}$ \\
~~Width (eV)\dotfill & 0$^{b}$ & 0$^{b}$ & 0$^{b}$ & 0$^{b}$ & 0$^{b}$ & 0$^{b}$ \\
~~Normalization$^{a}$\dotfill & 130$^{+20}_{-62}$ & 183$^{+20}_{-50}$ & 91$^{+31}_{-76}$ & 118$^{+18}_{-75}$ & 85$^{+9}_{-50}$ & 95$^{+11}_{-51}$ \\
\hline
O Ly$\beta$ \\
~~Center (eV)\dotfill & 775(774)$^{b}$ & 775(774)$^{b}$ & 774(776)$^{b}$ & 774(776)$^{b}$ & 774(781)$^{b}$ & 774(781)$^{b}$ \\
~~Width (eV)\dotfill & 0$^{b}$ & 0$^{b}$ & 0$^{b}$ & 0$^{b}$ & 0$^{b}$ & 0$^{b}$ \\
~~Normalization$^{a}$\dotfill & 40$^{+6}_{-7}$ & 76$\pm$10 & 12$^{+10}_{-11}$ & 29$\pm$14 & 18$\pm$8 & 29$^{+10}_{-9}$ \\
\hline
O Ly$\gamma$ \\
~~Center (eV)\dotfill & 817(816)$^{b}$ & 817(816)$^{b}$ & 816(818)$^{b}$ & 816(818)$^{b}$ & 816(823)$^{b}$ & 816(823)$^{b}$ \\
~~Width (eV)\dotfill & 0$^{b}$ & 0$^{b}$ & 0$^{b}$ & 0$^{b}$ & 0$^{b}$ & 0$^{b}$ \\
~~Normalization$^{a}$\dotfill & 0 ($<$5) & 36$^{+15}_{-24}$ & 0 ($<$10) & 11
($<$32) & 0 ($<$11) & 9 ($<$23) \\ 
\hline
\\
O Ly$\delta$ \\
~~Center (eV)\dotfill & 837(836)$^{b}$ & 837(836)$^{b}$ & 836(838)$^{b}$ & 836(838)$^{b}$ & 836(843)$^{b}$ & 836(843)$^{b}$ \\
~~Width (eV)\dotfill & 0$^{b}$ & 0$^{b}$ & 0$^{b}$ & 0$^{b}$ & 0$^{b}$ & 0$^{b}$ \\
~~Normalization$^{a}$\dotfill & 0 ($<$11) & 14 ($<$52) & 0 ($<$12) & 14 ($<$28)
& 6 ($<$18) & 11 ($<$66) \\
\hline
O Ly$\epsilon$ \\
~~Center (eV)\dotfill & 849(848)$^{b}$ & 849(848)$^{b}$ & 848(850)$^{b}$ & 848(850)$^{b}$ & 848(855)$^{b}$ & 848(855)$^{b}$ \\
~~Width (eV)\dotfill & 0$^{b}$ & 0$^{b}$ & 0$^{b}$ & 0$^{b}$ & 0$^{b}$ & 0$^{b}$ \\
~~Normalization$^{a}$\dotfill & 16$^{+4}_{-11}$ & 20 ($<$35) & 5 ($<$12) & 0
($<$17) & 8 ($<$17) & 10 ($<$23) \\ 
\hline
Ne He$\alpha$ \\
~~Center (eV)\dotfill & 913(913)$\pm$1 & 914(914)$\pm$1 & 914(916)$\pm$2 & 914(916)$^{+2}_{-1}$ & 918(925)$^{+3}_{-2}$ & 918(925)$^{+3}_{-2}$ \\
~~Width (eV)\dotfill & 13$\pm$2 & 14$\pm$2 & 6$^{+5}_{-7}$ & 8$^{+4}_{-9}$ & 0$^{+15}_{-0}$ & 10$^{+5}_{-10}$ \\
~~Normalization$^{a}$\dotfill & 181$\pm$5 & 221$\pm$5 & 150$\pm$8 & 156$^{+8}_{-9}$ & 57$\pm$4 & 61$^{+4}_{-5}$ \\
\hline
Ne Ly$\alpha$ \\
~~Center (eV)\dotfill & 1022(1021)$^{b}$ & 1022(1021)$^{b}$ & 1021(1023)$^{b}$ & 1021(1023)$^{b}$ & 1021(1028)$^{b}$ & 1021(1028)$^{b}$ \\
~~Width (eV)\dotfill & 0$^{b}$ & 0$^{b}$ & 0$^{b}$ & 0$^{b}$ & 0$^{b}$ & 0$^{b}$ \\
~~Normalization$^{a}$\dotfill & 3$\pm$2 & 10$\pm$2 & 2 ($<$5) & 4$\pm$3 & 0
($<$2) & 2$\pm$1 \\ 
\hline
Ne He$\beta$ \\
~~Center (eV)\dotfill & 1074(1073)$^{b}$ & 1074(1073)$^{b}$ & 1073(1075)$^{b}$ & 1073(1075)$^{b}$ & 1073(1080)$^{b}$ & 1073(1080)$^{b}$ \\
~~Width (eV)\dotfill & 0$^{b}$ & 0$^{b}$ & 0$^{b}$ & 0$^{b}$ & 0$^{b}$ & 0$^{b}$ \\
~~Normalization$^{a}$\dotfill & 10$\pm$2 & 13$\pm$2 & 7$^{+2}_{-3}$ & 8$\pm$3 & 2$^{+2}_{-1}$ & 3$\pm$2 \\
\hline
Ne He$\gamma$ \\
~~Center (eV)\dotfill & 1127(1126)$^{b}$ & 1127(1126)$^{b}$ & 1126(1128)$^{b}$ & 1126(1128)$^{b}$ & 1126(1133)$^{b}$ & 1126(1133)$^{b}$ \\
~~Width (eV)\dotfill & 0$^{b}$ & 0$^{b}$ & 0$^{b}$ & 0$^{b}$ & 0$^{b}$ & 0$^{b}$ \\
~~Normalization$^{a}$\dotfill & 3$^{+2}_{-1}$ & 6$^{+2}_{-1}$ & 4$\pm$2 &
5$\pm$2 & 0 ($<$4) & 1 ($<$3) \\
\hline
Ne He$\delta$ \\
~~Center (eV)\dotfill & 1150(1149)$^{b}$ & 1150(1149)$^{b}$ & 1149(1151)$^{b}$ & 1149(1151)$^{b}$ & 1149(1156)$^{b}$ & 1149(1156)$^{b}$ \\
~~Width (eV)\dotfill & 0$^{b}$ & 0$^{b}$ & 0$^{b}$ & 0$^{b}$ & 0$^{b}$ & 0$^{b}$ \\
~~Normalization$^{a}$\dotfill & 2$\pm$1 & 2$\pm$1 & 0 ($<$2) & 0 ($<$1) & 0
($<$2) & 1 ($<$4) \\ 
\hline
Ne Ly$\beta$ \\
~~Center (eV)\dotfill & 1210(1209)$^{b}$ & 1210(1209)$^{b}$ & 1209(1211)$^{b}$ & 1209(1211)$^{b}$ & 1209(1216)$^{b}$ & 1209(1216)$^{b}$ \\
~~Width (eV)\dotfill & 0$^{b}$ & 0$^{b}$ & 0$^{b}$ & 0$^{b}$ & 0$^{b}$ & 0$^{b}$ \\
~~Normalization$^{a}$\dotfill & 2$\pm$1 & 4$\pm$1 & 1 ($<$2) & 2$\pm$1 & 0
($<$1) & 0 ($<$1) \\ 
\hline
Mg He$\alpha$ \\
~~Center (eV)\dotfill & 1343(1343)$^{+7}_{-9}$ & 1342(1342)$^{+5}_{-6}$ & --- & --- & 1350(1357)$\pm10$ & 1346(1353)$\pm15$ \\
~~Width (eV)\dotfill & 0 ($<$16) & 13 ($<$24) & 0 ($<$103) & 0 ($<$75)
& 0 ($<$1796) & 17 ($<$42) \\
~~Normalization$^{a}$\dotfill & 3.1$\pm$1.1 & 5.5$\pm$1.2 & 1.3 ($<$2.9) &
2.0$^{+4.5}_{-1.3}$ & 1.5$^{+1.1}_{-1.0}$ & 2.1$^{+1.1}_{-1.0}$ \\ 
\hline
Mg He$\beta$ \\
~~Center (eV)\dotfill & 1575(1575)$\pm22$ & 1570(1570)$^{+18}_{-17}$ & 1538(1540)$^{+38}_{-36}$ & 1537(1540)$^{+33}_{-30}$ & --- & --- \\
~~Width (eV)\dotfill & 0$^{b}$ & 0$^{b}$ & 0$^{b}$ & 0$^{b}$ & 0$^{b}$ & 0$^{b}$ \\
~~Normalization$^{a}$\dotfill & 0.8 ($<$1.6) & 1.2$\pm$0.8 & 0.7 ($<$1.7) & 0.8
($<$1.8) & 0.1 ($<$0.9) & 0.2 ($<$1.0) \\
\hline
Si He$\alpha$ \\
~~Center (eV)\dotfill & 1835(1835)$\pm15$ & 1834(1834)$\pm15$ & --- & --- & 1866(1873)$\pm36$ & 1864(1871)$\pm34$ \\
~~Width (eV)\dotfill & 0$^{b}$ & 0$^{b}$ & 0$^{b}$ & 0$^{b}$ & 0$^{b}$ & 0$^{b}$ \\
~~Normalization$^{a}$\dotfill & 0.9$^{+0.9}_{-0.7}$ & 1.0$^{+1.0}_{-0.7}$ & 0.8
($<$1.9) & 0.8 ($<$1.9) & 1.0$\pm$0.8 & 1.1$\pm$0.8 \\
\enddata
\tablecomments{Columns A and B are responsible for the thermal
  emission models with $n_\mathrm{e}t =
  2.8\times10^{11}$\,cm$^{-3}$\,sec$^{-1}$ and $n_\mathrm{e}t =
  1.4\times10^{11}$\,cm$^{-3}$\,sec$^{-1}$, respectively.  Other notes
  can be found in Table~\ref{tab:bremss_gauss}.
} 
\label{tab:cx}
\end{deluxetable}

%\begin{deluxetable}{lcc|cc|cc}
%\tabletypesize{\scriptsize}
%\tablecaption{Ion fractions in the thermal plasma$^a$.}
%\tablewidth{0pt}
%\tablehead{
%\colhead{Ion}&\multicolumn{2}{c}{Fraction at $n_{\rm
%    e}t$=5$\times10^{10}$}&\multicolumn{2}{c}{Fraction at $n_{\rm
%    e}t$=1$\times10^{11}$}&\multicolumn{2}{c}{Fraction at $n_{\rm
%    e}t$=2$\times10^{11}$}} 
%\startdata
%& From K$\alpha$ & From K$\beta$ & From K$\alpha$ & From K$\beta$ &
%From K$\alpha$ & From K$\beta$ \\ 
%O VIII\dotfill & 0.095 & 0.091 & 0.18 & 0.18 & 0.30 & 0.29 \\
%O IX\dotfill & 0.033 & 0 & 0.031 & 0 & 0.036 & 0\\
%Ne X\dotfill & 0.004 & 0.003 & 0.009 & 0.008 & 0.017 & 0.016 \\
%\enddata
%\tablecomments{$^a$$kT_{\mathrm e}$ = 0.2\,keV and the abundance ratio
%  of Ne/O = 2 times the solar value.} 
%
%\label{tab:ion_frac}
%\end{deluxetable}

%\end{document}

\end{document}